\documentstyle[aps,prd,graphicx,amsfonts,epsf,bbm]{revtex}  

\newcommand{\beq}{\begin{equation}}
\newcommand{\eeq}{\end{equation}}

\begin{document}
\preprint{hep-ph/0004067}

\draft

\title{R-Parity Violation and the Decay $b\rightarrow s \gamma$}
\author{Th. Besmer and A. Steffen}
\address{Institute of Theoretical Physics, University of Z\"urich, Switzerland}
\date{\today}

\maketitle

\begin{abstract}

We investigate the influence of the R-parity violating couplings
$\lambda$, $\lambda'$ and $\lambda''$ on the branching ratio of
$b\rightarrow s\gamma$ in leading logarithmic approximation. The 
operator
basis is enlarged and the corresponding $\gamma$-matrix calculated. 
The
matching conditions receive new contributions from the R-parity 
violating sector. The comparison with the experiment is rather difficult due to
the model dependence of the result.
\end{abstract}


\section{Introduction}

The decay $b\rightarrow s\gamma$, forbidden at tree-level in the Standard Model 
(SM),
is an excellent candidate for exploring the influence of new physics
beyond the SM. However, the experimentally measured branching ratio
$\mbox{Br}(B\rightarrow X_s\gamma)=(3.15\pm0.93)\cdot 10^{-4}$ \cite{exp} is in perfect
agreement with the SM prediction computed at the next-to-leading
order:$\mbox{Br}(b\rightarrow s\gamma)_{SM}=(3.28\pm 0.30)\cdot 10^{-4}$ \cite{theor}. 
This
leads to the conclusion that the influence of new physics on this 
decay
is either very limited or the new  contributions cancel among each 
other
to a large extent. The most serious and attractive extention of the Standard Model is
Supersymmetry (SUSY). It has a variety of very appealing features. For
instance it provides a natural solution to the hierarchy problem. SUSY
doubles the particle spectrum of the SM providing every fermion with a
bosonic partner and vice versa. If SUSY were an exact symmetry of 
nature
the new particles would be of the same mass as their partners. This is
definitely excluded from
 experiment. Therefore SUSY must be broken. To retain the solution of 
the
hierarchy problem one allows only a breaking which does not introduce
quadratic divergences in loop diagrams. Although this reduces the set 
of
possible breaking terms substantially to the so-called
\emph{soft-breaking} terms the number of free parameters of softly
broken SUSY still exceeds one hundred. This theory, i.e., SUSY with soft
breaking terms \emph{and no others}, emerges naturally as
low energy limit of local supersymmetry, \emph{Supergravity
(SUGRA)} by breaking it at some high scale $\sim 10^{11}$ GeV and 
taking the so-called\emph{ flat limit} \cite{nilles}. The connection to
supergravity eliminates much of the freedom in choosing the parameters
for the soft breaking terms, enhancing the predictive power of the
model.\\ 
Viewing naturalness as a first principle in model building one
runs immediately into a problem: 
The Yukawa interactions are fixed by the so-called 
\emph{Superpotential
$W$}, which is the most general third degree polynomial that can be 
built 
of gauge invariant combinations of the (left-handed) superfields of 
the theory. In the case of the minimal extension of the standard 
model with gauge group $SU(3)\times SU(2)\times U(1)$ it shows the 
form
\begin{eqnarray}
W&=&\lambda^{u}_{ij}Q_{i}H_{2}U_{j}^{c}+\lambda^{d}_{ij}H_{1}Q_{i}D_{j}^{c}
+\lambda^{e}_{ij}H_{1}L_{i}E_{j}^{c}+\mu
H_{1}H_{2} \nonumber \\
&&+\mu_{2i}L_{i}H_{2}+\frac{1}{2}\lambda_{ijk}L_{i}L_{j}E_{k}^{c}+
\lambda'_{ijk}L_{i}Q_{j}D_{k}^{c}+\frac{1}{2}
\lambda''_{ijk}U_{i}^{c}D_{j}^{c}D_{k}^{c},
\label{W1}
\end{eqnarray}
where $Q,\,U^{c},\,D^{c},\,L,\,E^{c},\,H_{1}$ and $H_{2}$ label the
left-handed superfields that describe left- and right-handed (s)quarks
and (s)leptons and the Higgs-bosons (-fermions) respectively.
The terms on the second line of (\ref{W1}) lead to unwanted baryon- 
and
lepton-number violating vertices. The product
$\lambda'_{ijk}\cdot\lambda''_{ijk}$, for instance, is restricted to 
be
smaller than
$10^{-10}$ \cite{smirnov}.  It is not obvious why the coefficients of these terms 
should
not be of order unity if the corresponding interaction is not 
protected by any symmetry. Hence one way to avoid these unwanted 
couplings is the invention of a new discrete
symmetry, called \emph{R-Parity} \cite{dreiner1}. The multiplicative
quantum number $R$ is then defined as
\beq
R=(-1)^{3B+L+2s},
\eeq
where $s$ is the spin of the particle.
The particles of the standard model are then R-parity even fields 
while their 
supersymmetric partners are R-parity odd fields. The superfields 
adopt the value
of R from their scalar components. The terms on the 
second line of (\ref{W1}) are then forbidden by this symmetry. One 
ends up with the \emph{Minimal Supersymmetric Standard Model (MSSM)}
\cite{mssm1,mssm2,mssm3}. 
One could even go one step further and promote this new symmetry to a 
$U(1)$ 
gauge symmetry (\emph {R-symmetry}). The R-charges must then be 
chosen such that the ``nice'' terms 
remain in the Lagrangian and the unwanted terms will not be allowed 
anymore. Further requirement is the vanishing of possible 
additional anomalies \cite{chams1,chams2}. This scenario is preferred from a string 
theoretical view because in string theory one has many additional 
$U(1)$'s ``floating around'', what makes the 
introduction of this symmetry more natural.\\
It is interesting to explore what the constraints on the R-parity 
violating couplings, especially $\lambda_{ijk}$, $\lambda'_{ijk}$, and 
$\lambda''_{ijk}$, are 
from the experimental point of view. Bounds on these couplings have 
been found by many authors \cite{bounds1,bounds2,bounds3,bounds4}. Sometimes the bounds on products of
couplings 
are more restrictive than the products of the individual bounds. 
Depending on the reactions that the couplings are involved in the 
constraints from experiment are very strong or rather poor.\\
In this paper we want to explore the (theoretical) influence of 
$\lambda$, 
$\lambda'$, and $\lambda''$ on the decay $b\rightarrow s\gamma$. This 
has been done before by other authors \cite{decarlos1}. However, here we will include 
the full operator basis at leading-log of the effective low-energy 
theory.\\
The comparison with the experiment must result in bounds that are 
model 
dependent. The relevant couplings are proportional to the inverse 
mass-squared of particles (Higgs, SUSY-partners) that have not yet 
been 
detected. However, it is clear from next-to-leading-log calculations 
that the prediction of the MSSM with a realistic mass spectrum lies 
within the current experimental bounds.\\
Because of the strong model 
dependence we do not perform our calculations at 
highest precision. Nevertheless we try to include all the possibly 
relevant 
contributions at leading-log. We do not claim our results to be very 
accurate. 
It is the aim of this paper to explore if $b\rightarrow s\gamma$ has 
the potential power to reduce the bounds of (products of) some of the 
R-parity 
breaking couplings substantially, i.e., by some order of magnitude.\\
This article is divided as follows: Section 2 introduces the model we 
are working with. We try to describe as precise as possible what 
our assumptions are. The following section deals with the effective 
Hamiltonian approach. The enhanced operator basis is presented and 
the 
$\gamma$-matrix as well as the matching conditions at $M_{W}$ 
calculated. The 
comparison with the experiment is performed in section 4. Section 5 
contains our conclusions. In the appendix we present some technical 
details of our computations, namely the mixing matrices,the relevant part of the
interaction Lagrangian and the RGE's 
that are needed.

\section{Framework}

In supersymmetry the matter fields are described by left-handed 
chiral 
superfields $\Sigma^{i}$. They contain a scalar boson $z^{i}$ and a 
two-component fermion $\psi^{i}$. Real vector superfields $V^{a}$ are 
needed to form the gauge bosons $A^{a}_{\mu}$ and the gauginos 
$\lambda^{a}$. The minimal supersymmetric standard model is the model 
with the smallest particle content that is able to mimic the features 
of the standard model, i.e., including all the observed particles, 
gauge group $SU(3)_{\mathrm{colour}}\times 
SU(2)_{\mathrm{weak}}\times U(1)_{Y}$, spontaneous symmetry breaking 
and the 
Higgs mechanism. Its superfields together with their components are 
collected in table \ref{tbl:MSSM}.
A few comments are in order:
\begin{itemize}
    \item
    Because the theory only deals with left-handed chiral superfields 
    the $SU(2)$-singlet matter fields must be defined via their 
charged 
    conjugated (anti-)\ fields.
    \item
    No conjugated superfields are allowed in the superpotential 
    $W$. Therefore we need to introduce a second Higgs field to give 
    the up and the down quarks a mass when the neutral components 
    gain a vacuum expectation value (vev).
    \item
    The $L$'s and $R$'s in the names of squarks and sleptons only 
    identify the fermionic partners. These fields are just normal 
    complex scalar bosons.
\end{itemize}
It must be mentioned that the fields in table \ref{tbl:MSSM} are 
\emph{not} the physical 
fields. 
\begin{itemize}
    \item
    In the Higgs sector three degrees of freedom 
    are eaten by the gauge bosons in analogy to the SM. We end up 
with one charged and three 
    neutral Higgs bosons.
    \item
    Higgsinos and gauginos of $SU(2)\times U(1)$ mix to form 
    \emph{charginos} and \emph{neutralinos}.
    \item
    Photon, $W$- and $Z$-boson form when the electroweak symmetry 
breaks down.
    \item
    The three generations of quarks and leptons mix via the 
    Cabibbo-Koba\-yashi-Maskawa-matrix $K$ to give the 
    mass eigenstates in complete analogy to the standard model.
    \item 
    The family mixing also takes place in the squark and slepton 
sector. 
    However, there is an 
    additional mixing between the partners of left- and right-handed 
fermions due to the soft-breaking terms.
\end{itemize}
Appendix \ref{ami} gives a more detailed description of the different 
mixings 
including a complete listing of all relevant mixing matrices.
\vskip 0.8cm

The component expression of the Lagrangian we base our 
model on can be written as
\beq
{\mathcal L}={\mathcal L}_{\mathrm{kin}}+{\mathcal L}_{\mathrm{int}}+{\mathcal L}_{W}-V_{\mathrm{soft}},
\eeq
where 
\begin{itemize}
\item ${\mathcal L}_{\mathrm{kin}}$ and ${\mathcal L}_{\mathrm{int}}$ 
stand for
the kinetic energies, the interactions between chiral and gauge 
fields and part of the scalar potential.
\item 
${\mathcal L}_{W}$ contains the rest of the scalar
potential and the Yukawa interactions:
\beq
{\mathcal L}_{W}=-\sum_{i}\left|\frac{\partial W}{\partial 
z_{i}}\right|^{2}-\frac{1}{2}\sum_{ij}\left[\psi_{i}\frac{\partial^{2}W}{\partial
z_{i}\partial z_{j}}\psi_{j}+\mbox{h.c.}\right],
\eeq
where $W$ should be viewed as a function of the scalar fields.
\item $W$ is the superpotential which contains all possible gauge 
invariant combinations of the left-handed superfields (but not their 
conjugated right-handed partners). In our case, this results in 
equation (\ref {W1}).
It should be noted that every term is a gauge 
invariant combination of the corresponding superfields, for 
instance, $Q_{i}H_{2}U_{j}^{c}$ is abbreviated for 
$\epsilon_{\alpha\beta}\,Q_{i}^{\alpha 
A}\,H_{2}^{\beta}\,U_{j}^{cA}$, where 
$\alpha$ and $\beta$ are $SU(2)$-indices, $A$ is an $SU(3)$-index 
and $\epsilon_{\alpha\beta}$ is the completely antisymmetric tensor 
with
$\epsilon_{12}=1$ \footnote{The antisymmetry of the $SU(2)$ product is the reason
why a term $\sim H_1H_1E_i^c$ is not introduced}. The second line of (\ref{W1}) 
represents the R-parity breaking sector. As a consequence of the 
antisymmetry in
the fields $\lambda$ is antisymmetric in its first two indices and 
$\lambda''$
is antisymmetric in its last two indices. Therefore the trilinear 
R-parity breaking
couplings of $W$ contain $9+27+9=45$ new parameters. 
\item $V_{\mathrm{soft}}$ includes the soft-breaking trilinear terms 
of the scalar
potential and mass terms for the scalar fields and the gauginos. It 
has the form
\begin{eqnarray}
V_{\mathrm{soft}}&=&[h^{u}_{ij}\tilde{q}_{i}\tilde h_{2}\tilde{u}_{j}^{\dagger}
+h^{d}_{ij}\tilde h_{1}\tilde{q}_{i}\tilde{d}_{j}^{\dagger}
+h^{e}_{ij}\tilde h_{1}\tilde{\ell}_{i}\tilde{e}_{j}^{\dagger} \nonumber \\
&&+\frac{1}{2}C_{ijk}\tilde{\ell}_{i}\tilde{\ell}_{j}\tilde{e}^{\dagger}_{k}
+C'_{ijk}\tilde{\ell}_{i}\tilde{q}_{j}\tilde{d}^{\dagger}_{k}
+\frac{1}{2}C''_{ijk}\tilde{u}^{\dagger}_{i}\tilde{d}^{\dagger}_{j}\tilde{d}^{\dagger}_{k}
\nonumber\\
&&+\tilde{\mu}\tilde{h}_{1}\tilde{h}_{2}+\tilde{\mu}_{2i}\tilde{\ell}_{i}\tilde{h}_{2}]+\mbox{h.c.}
+m_{ab}^2z^{a}z^{b\dagger} \nonumber \\
&&+\frac{1}{2}M_{a}[\lambda_{a}\lambda_{a}+\mbox{h.c.}].
\end{eqnarray}
\end{itemize}

\vskip 0.8cm

The MSSM in its full generality with an additional R-parity breaking 
sector 
involves over 150 
free parameters. These are by far too many for the model to be 
predictive. In
the following we will reduce the parameter space substantially by 
making some
assumptions that are, hopefully, well motivated.\\
As a first step it is important to mention that we see our
softly broken global SUSY at a low energy scale $\sim M_{Z}$ emerging 
from a
spontaneously broken local supersymmetry at a high scale $M_{X}\sim 10^{16}$
GeV taking the flat limit
$M_{\mathrm{Planck}}\rightarrow \infty$,
$m_{0}:=m_{\mathrm{gravitino}}=$constant. This fixes most of the 
parameters of
$V_{\mathrm{soft}}$\ at
$M_{X}$: \\
\begin{itemize}
\item
All the coefficients of the trilinear terms in $V_{\mathrm{soft}}$ 
are related
to the corresponding terms of $W$ by a multiplication with a 
universal factor
$Am_{0}$:

\beq
\begin{array}{lcl}
h^{u}_{ij}=Am_{0}\lambda^{u}_{ij}\hskip 1cm
&C_{ijk}=Am_{0}\lambda_{ijk}\\ \\
h^{d}_{ij}=Am_{0}\lambda^{d}_{ij}\hskip 1cm
&C'_{ijk}=Am_{0}\lambda'_{ijk}\\ \\
h^{e}_{ij}=Am_{0}\lambda^{e}_{ij}\hskip 1cm
&C''_{ijk}=Am_{0}\lambda''_{ijk}
\end{array}
\eeq

\item
An analogous statement holds for the bilinear terms:
\begin{eqnarray}
\tilde{\mu}=Bm_{0}\mu \hskip 1cm& \tilde{\mu}_{2i}=Bm_{0}\mu_{2i},
\end{eqnarray}
where usually
\beq
B=A-1.
\eeq

\item
The mass term of the scalars are diagonal and universally equal to 
the gravitino
mass $m_{0}$:
\beq
m_{ab}^2=m_{0}^2\delta_{ab}
\eeq

\end{itemize}

We assume unification of the gauge group at $M_{X}$. As a consequence 
all
the gaugino masses are equal at that scale:
\beq
M_{i}(M_X)=M\hskip 1cm\forall i
\eeq

Not all entries of the Yukawa-matrices $\lambda^{u}_{ij}$, 
$\lambda^{d}_{ij}$,
and $\lambda^{e}_{ij}$ are observable in the SM. One usually
chooses two of them (in most cases $\lambda^{d}$ and $\lambda^{e}$) 
to be diagonal. Although this is in principle not possible
in our model we will adopt this choice here for convenience. All the 
entries
at \emph{$M_{W}$}
are then fixed by the quark/lepton masses, the vevs $v_{1}$ and 
$v_{2}$ of the
neutral Higgs bosons $H_1$ and $H_2$ respectively and the Cabibbo-Kobayashi-Maskawa-matrix $K$.\\

$\mu H_{1}H_{2}$ is the so-called $\mu$-term. The mass parameter 
$\mu$ must be
of order of the weak scale whereas the natural scale would be the 
Planck
mass $M_{P}\sim 10^{19}$ GeV. The question why this parameter 
is so small is referred to as the  \emph{$\mu$-problem}.\\

$\mu_{2i}L_{i}H_{2}$ and 
$\tilde{\mu}_{2i}\tilde{\ell}_{i}\tilde{h}_{2}$ mix Higgs and 
leptonic sector. We choose to
set $\mu_{2i}(M_X)=\tilde{\mu}_{2i}(M_{X})=0$.  At $M_{Z}$,
$\mu_{2i}L_{i}H_{2}$ can be rotated away with the help of a field
redefinition of the Higgs field whereas
$\tilde{\mu}_{2i}$ is, at least in the case of a physically realistic spectrum,
 small enough to be
neglected.\\ One ends up with the following free parameters:
\beq
A,\; m_{0},\; M,\; \mu
\eeq
Usually, one replaces one of these parameters by 
$\tan\beta=v_2/v_1$.
A second parameter will be fixed by the requirement of a correct
electroweak symmetry breaking. This means, the minimum of the scalar
Higgs potential must occur at values $(v_{1},v_{2})$ which reproduce 
the
correct mass of the
$Z$-boson:
\beq
M_{Z}^2=\frac{g^2_{1}+g^2_{2}}{2}(v_{1}^{2}+v_{2}^{2})
\eeq
It has been realized by many authors \cite{gamberini1,gamberini2,gamberini3} that the tree-level potential 
\begin{eqnarray}
V_{0}&=&(\mu^2+m^2_{H_1H_1}) (\tilde h^0_1)^2 +(\mu^2+m^2_{H_2H_2}) 
(\tilde h^0_2)^2 +2 \tilde\mu\,\tilde h^0_1\,\tilde h^0_2\\ \nonumber
&&+\frac{g_1^2+g_2^2}{8}
\left[(\tilde h^0_1)^2-(\tilde h^0_2)^2\right]^2
\end{eqnarray}
is
not enough to gain sensible values for $v_{1}$ and $v_{2}$. Thus we 
have to
include the first correction $\Delta V$ to the effective potential. 
At a
mass scale $Q$, it has the form \cite{coleman1,guitsos}
\begin{eqnarray}
\Delta V & = & \frac{1}{64
\pi^{2}}\mbox{Str}\left[{\mathcal M}^{4}\left(\ln{\frac{{\mathcal M}^{2}}{Q^{2}}}-\frac{3}{2}\right)\right]
\nonumber \\
& = & \frac{1}{64 \pi^{2}}\sum_{p}(-1)^{2s_{p}}
(2s_{p}+1)n_{p}M_{p}^{4}\left(\ln{\frac{|M_{p}^{2}|}{Q^{2}}}-\frac{3}{2}\right).
\end{eqnarray}
Here, Str denotes the supertrace and ${\mathcal M}^{2}$ is the 
tree-level mass
matrix squared. $p$ runs over all particles of the theory with spin $s_p$ 
whereas $M_{p}$ is the
corresponding eigenvalue (mass) of the particle. $n_{p}$ counts for 
the degrees
of freedom according to colour and helicity. The eigenvalues $M_{p}$ 
depend on
the neutral components of $\tilde h_{1}$ and $\tilde h_{2}$ and therefore change 
the shape of
the potential and its minimum. For most of the particles, they can 
only be
computed numerically. One comment is in order: A phase rotation of the
Higgs fields can turn a negative vev into a positive one. This freedom
is reflected in the fact that the \emph{sign} of $\mu$ can be chosen
freely giving then a different phenomenology.\\ To compute the mixing
matrices and the one-loop effective potential one has to know the mass
matrices at
$M_{Z}$. Unfortunately, for most of the parameters we know the 
boundary
conditions at the high scale $M_{X}$. Hence one has to set up the
complete set of RGE's to run the parameters from
$M_{X}$ to
$M_{Z}$. The details of how to get
a consistent parameter space are described in section \ref{results}, 
the complete set of RGE's can be found in appendix \ref{ar}.

\section{The effective Hamiltonian}
\subsection{The case of the MSSM}

\begin{figure}[btp]

\begin{center}
\epsfbox{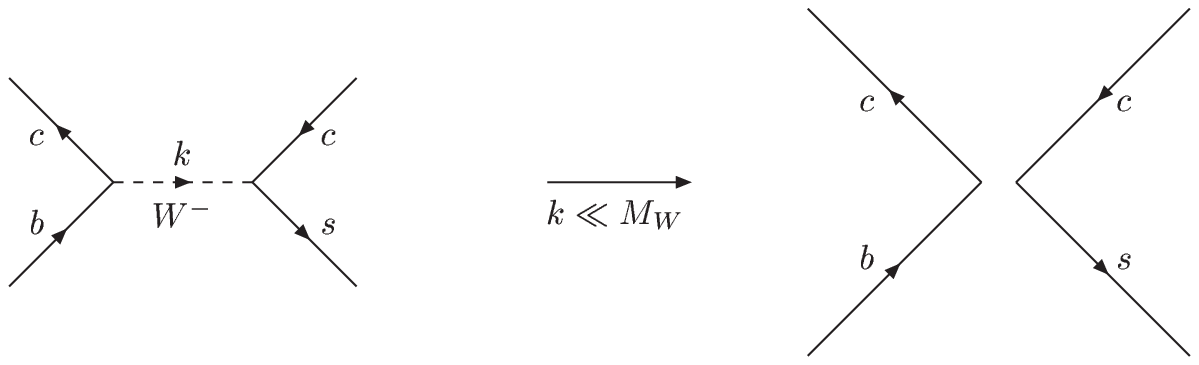}
\end{center}
\caption{\label{fig:1} From the full to the effective theory}
\end{figure}

The decay $b\rightarrow s\gamma$ occurs at energies of a few GeV$\sim
m_{b}$. This is
much below the weak scale. It makes sense to work with an effective 
Hamiltonian
${\mathcal H}_{{\mathrm{eff}}}$
where all the heavy fields (compared to $m_{b}$) are integrated out 
\cite{buras1}. 
In
the SM these are the $Z$- and the $W$-boson and the top quark whereas 
for
our purposes the $Z$-boson does not play any role. A result of
integrating out these fields is the appearance of new local operators 
of
dimension higher than four. This can be illustrated by the shrinking 
of
the Feynman diagrams in Fig.\ \ref{fig:1}. \\
Once one lets the strong interaction come into the game,
 QCD corrections of these new operators give rise to additional
operators. For a
 consistent treatment of all the corrections at a certain level of 
QCD we have
 to include a set of operators $O_{i}$ which closes under these
corrections. Our
 low-energy theory is then described by an effective Hamiltonian
 \beq
 {\mathcal H}_{\mathrm{eff}}=-\sum_{i}C_{i}O_{i}.
 \eeq
 
The QCD renormalization of the operators must be performed at a 
scale where no
 large logarithms appear, i.e., at $\mu\sim M_{W}$. However, a calculation of 
$\mbox{Br}(b\rightarrow
s\gamma)$ at energy scales $\sim m_{b}$
 requires the knowledge of the Wilson coefficients $C_{i}$ at 
\emph{that}
scale. The $C_{i}$'s depend on a renormalization scale $\mu$. They 
obey the renormalization group equations
\begin{eqnarray}
\frac{dC_{i}}{d\ln \mu}=\gamma_{ji}C_{j},
\end{eqnarray}
where $\gamma_{ij}$ is the gamma-matrix that emerges from the QCD-mixing of 
the
operators $O_{i}$. As initial conditions we find $C_{i}(M_{W})$ by 
matching the
effective theory with the full theory at that scale. The RGE's are 
then solved to
give $C_{i}(m_{b})$\\

In the standard model the relevant set of operators for 
$b\rightarrow s\gamma$
 is given by
\begin{eqnarray}
 O_{1} & = & 
(\overline{s_L}_\alpha\gamma^{\mu}b_{L\alpha})\;(\overline{c_L}_\beta\gamma_{\mu}c_{L\beta}) 
\nonumber \\ \nonumber \\
 O_{2} & = & 
(\overline{s_L}_\alpha\gamma^{\mu}b_{L\beta})\;(\overline{c_L}_\beta\gamma_{\mu}c_{L\alpha}) 
\nonumber \\ \nonumber \\
 O_{3} & = & 
(\overline{s_L}_\alpha\gamma^{\mu}b_{L\alpha})\;\sum_{i}(\overline{
q_L}_{i\beta}\gamma_{\mu}q_{Li\beta}) 
\nonumber \\ \nonumber \\
 O_{4} & = &
(\overline{s_L}_\alpha\gamma^{\mu}b_{L\beta})\;\sum_{i}(\overline{q_L}_{i\beta}\gamma_{\mu}q_{Li\alpha})
\nonumber \\ \nonumber \\
 O_{5} & = &
(\overline{s_L}_\alpha\gamma^{\mu}b_{L\alpha})\;\sum_{i}(\overline{q_R}_{i\beta}\gamma_{\mu}q_{Ri\beta}) 
\nonumber \\\nonumber \\
 O_{6} & = &
(\overline{s_L}_\alpha\gamma^{\mu}b_{L\beta})\;\sum_{i}(\overline{q_R}_{i\beta}\gamma_{\mu}q_{Ri\alpha})
\nonumber \\ \nonumber \\
 O_{7} & = & \frac{e}{16 
\pi^2}\;m_{b}\;\overline{s_L}_\alpha\sigma_{\mu\nu}b_{R\alpha}\;F^{\mu\nu} 
\nonumber \\ \nonumber \\
 O_{8} & = & \frac{g_{s}}{16 
\pi^2}\;m_{b}\;\overline{s_L}_\alpha\sigma_{\mu\nu}b_{R\beta}\;t^{a\alpha\beta}\;G^{a\mu\nu}. 
 \label{basis1}
 \end{eqnarray}

\begin{figure}[tp]

\begin{center}
\epsfbox{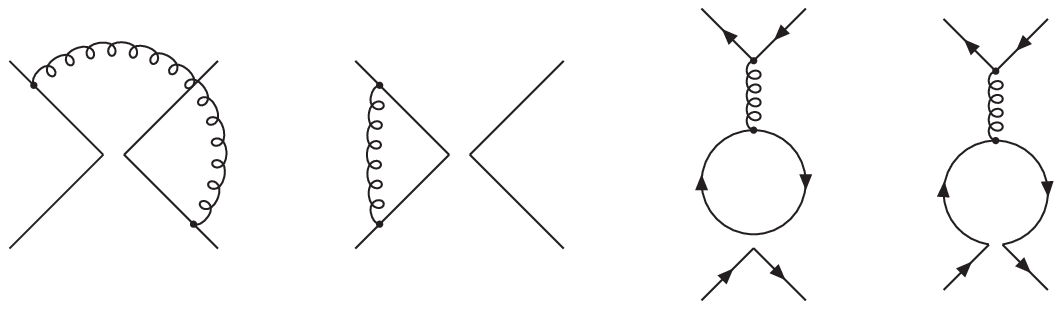}
\end{center}
\caption{\label{fig:2} Diagrams contributing to the 1-loop mixing of 
the
four-fermion ope\-rators.}
\end{figure}

Here, $\overline{q_{L/R}}\gamma^{\mu}q_{L/R}=\overline
q\gamma^{\mu}(1\mp\gamma^{5})q$, $\overline 
s_{L}\sigma_{\mu\nu}b_{R}=\overline
s\sigma_{\mu\nu}(1+\gamma^{5})b$, the sum runs over the five quarks 
of the
effective theory at $m_{b}$, and $\alpha$, $\beta$ are colour indices. 
The
operators $O_{1}$ and $O_{3}-O_{6}$ are introduced by QCD corrections 
through diagrams
like those depicted in Fig.\ \ref{fig:2}.\\

\begin{figure}[bp]

\begin{center}
\epsfbox{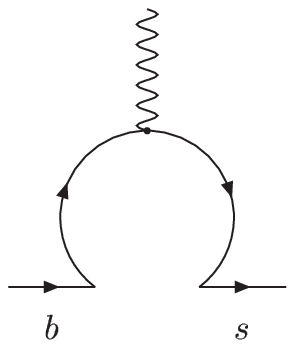}
\end{center}
\caption{\label{fig:3} Diagram that gives finite contributions.}
\end{figure}

It is interesting to note that in the MSSM without 
R-parity
violation the relevant operator basis does not change although 
we definitely
have new decay channels. The $\gamma$-matrix at leading-log can be 
found in
appendix \ref{agamma} by picking up the relevant entries. 
Because the
contribution of the diagram in Fig.\ \ref{fig:3} is not divergent,
 the mixing of $O_{1}-O_{6}$ with $O_{7/8}$ 
 involves two-loop diagrams like those of figures  
\ref{fig:4} and
\ref{fig:5} \cite{grinstein1,grinstein2}.

Although we only need the divergent part of these diagrams one has to 
be very
careful in computing the counterterms because one must include certain
additional operators \cite{herrlich,buras2}, so-called \emph{evanescent operators} that 
vanish in four
dimensions but must be kept in $D$ dimensions in all intermediate steps of the
calculation. Furthermore, the two-loop results 
are regularization scheme dependent \cite{buras2}. This regularization scheme 
dependence of
the $\gamma$-matrix cancels with possible finite one-loop [but $O(\alpha_s^0)$] 
contributions from $O_{5}$
and $O_{6}$, inserted in the diagram of Fig.\ \ref{fig:3}, to the matrix element
of $b\rightarrow s\gamma$. As a 
result of these complications, the
calculation of the $\gamma$-matrix at leading-log has been finished 
only few
years ago \cite{ciuchini1,ciuchini2,misiak1,misiak2}. Meanwhile, also the next-to leading result
is known
\cite{nexttoleading1,nexttoleading2,nexttoleading3,nexttoleading4,nexttoleading5,nexttoleading6,nexttoleading7,nexttoleading8,nexttoleading9,nexttoleading10,nexttoleading11,nexttoleading12,nexttoleading13,nexttoleading14,buras2}. 
We 
will
concentrate on the leading-log calculation.\\

The matching of the effective theory with the full theory at $M_{W}$ 
must be
performed only at order $\alpha_{s}^{0}$ because the leading QCD 
corrections are
already included in the operator mixing. In the standard model this 
involves
diagrams with a $W$-exchange for $O_{2}$ and diagrams with a 
$W$-$t$-loop (in
the unitary gauge) for
$O_{7/8}$ corresponding to Fig.\ \ref{fig:6} (a). The result is then \cite{inami}

\begin{eqnarray}
C_{2\mathrm{SM}}(M_{W}) & = & \frac{G_{F}}{\sqrt{2}}K_{ts}^{*}K_{tb} 
\\
C_{7\mathrm{SM}}(M_{W}) & = & -\frac{G_{F}}{\sqrt{2}}K_{ts}^{*}K_{tb} 
3 x_{tW}
[Q_{u}F_{1}(x_{tW})+F_{2}(x_{tW})] \\
C_{8\mathrm{SM}}(M_{W}) & = & -\frac{G_{F}}{\sqrt{2}}K_{ts}^{*}K_{tb} 
3 x_{tW}
F_{1}(x_{tW}),
\end{eqnarray}
where $G_{F}=g_{2}^2/(4\sqrt{2}M_{W}^2)\approx 
1.166\cdot10^{-5}$
GeV$^{-2}$ is the Fermi constant, $x_{ab}=m_{a}^2/m_{b}^{2}$ 
and the functions $F_{i}$ are
given in appendix \ref{af}.\\

The matching conditions for $C_{7/8}$ become far more complicated in 
the case of
the MSSM, even without R-parity violation. In addition to the 
$W$-$t$-loop there are four more combinations of particles in the loop: 
\begin{itemize}
\item
charged Higgs $H^{\pm}$---top $t$ ,Fig.\ \ref{fig:6} (b)
\item
up-squark $\tilde u$---chargino $\chi^{\mathrm{ch}}$ ,Fig.\ \ref{fig:6} (c)
\item 
down-squark $\tilde d$---neutralino $\chi^{0}$ ,Fig.\ \ref{fig:6} (d)
\item
down-squark $\tilde d$---gluino $g$ ,Fig.\ \ref{fig:6} (e)
\end{itemize}

In principle, these contributions give rise to two new operators 

\begin{eqnarray}
\tilde O_{7} & = & \frac{e}{16 
\pi^2}\;m_{b}\;\overline{s_R}_\alpha\sigma_{\mu\nu}b_{L\alpha}\;F^{\mu\nu} 
\nonumber \\ \nonumber \\
 \tilde O_{8} & = & \frac{g_{s}}{16 
\pi^2}\;m_{b}\;\overline{s_R}_\alpha\sigma_{\mu\nu}b_{L\beta}\;t^{a\alpha\beta}\;G^{a\mu\nu}. 
 \end{eqnarray}
 
 Note that $\tilde O_{7/8}$ differ from $O_{7/8}$ only by their
 handedness. The Wilson coefficients of these new operators are usually
 so small that they can be safely neglected \cite{bertolini}. However, we will include 
them in
 our calculations because we need them later on anyway. We only 
neglect
 contributions from the Higgs sector which are proportional to some 
light
quark
 masses.\\
 The matching conditions become rather involved now. They include
 many mixing matrices, whose definitions we give in appendix 
\ref{am}. We have \cite{bertolini}
 
 \begin{eqnarray}
 C_{7\mathrm{MSSM}} & = & C_{7\mathrm{SM}} \nonumber \\
 &&- \frac{G_{F}}{\sqrt{2}}K_{ts}^{*}K_{tb}\{\cot^2\beta\,
 x_{tH}[Q_{u}F_1(x_{tH})+F_2(x_{tH})] \nonumber \\
 &&\phantom{- \frac{G_{F}}{\sqrt{2}}K_{ts}^{*}K_{tb}\{}
 +x_{tH}[Q_uF_3(x_{tH})+F_4(x_{tH})]\}
 \nonumber \\
 &&+\frac{1}{4}\frac{1}{m^2_{\tilde
 u_j}}B^d_{2j\ell}B^{d*}_{3j\ell}\left[F_1(x_{\chi^{\mathrm{ch}}_\ell \tilde
u_j})+Q_{u}F_2(x_{\chi^{\mathrm{ch}}_\ell \tilde
 u_j})\right] \nonumber\\
&&+\frac{1}{4}\frac{1}{m^2_{\tilde
 u_j}}\frac{m_{\chi^{\mathrm{ch}}_\ell}}{m_b}B^d_{2j\ell}A^{d*}_{3j\ell}\left[F_3(x_{\chi^{\mathrm{ch}}_\ell
 \tilde u_j})+Q_u F_4(x_{\chi^{\mathrm{ch}}_\ell
 \tilde u_j})\right] \nonumber \\
  &&+ \frac{Q_d}{4}\frac{1}{m^2_{\tilde
 d_j}}\left[D^d_{2j\ell}D^{d*}_{3j\ell}F_2(x_{\chi^0_\ell \tilde
 d_j})+\frac{m_{\chi^0_\ell}}{m_b}D^d_{2j\ell} 
C^{d*}_{3j\ell}F_4(x_{\chi^0_\ell
 \tilde d_j})\right] \nonumber \\
 &&+ \frac{2}{3}Q_d\,g_s^2\frac{1}{m^2_{\tilde
 d_j}}\left[\Gamma^{d\dagger}_{L2j}\Gamma^d_{Lj3}F_2(x_{g\tilde
 d_j})-\frac{m_g}{m_b}\Gamma^{d\dagger}_{L2j}\Gamma^d_{Rj3}F_4(x_{g\tilde
 d_j})\right] 
 \\
 \nonumber \\
 C_{8\mathrm{MSSM}} & = & C_{8\mathrm{SM}} \nonumber \\
 &&- \frac{G_{F}}{\sqrt{2}}K_{ts}^{*}K_{tb}[\cot^2\beta\,
x_{tH}F_1(x_{tH})+x_{tH}F_3(x_{tH})] \nonumber \\
&&+ \frac{1}{4}\frac{1}{m^2_{\tilde
u_j}}\left[B^d_{2j\ell}B^{d*}_{3j\ell}F_2(x_{\chi^{\mathrm{ch}}_\ell \tilde
u_j})+\frac{m_{\chi^{\mathrm{ch}}_\ell}}{m_b}B^d_{2j\ell}A^{d*}_{3j\ell}F_4(x_{\chi^{\mathrm{ch}}_\ell
\tilde u_j})\right] \nonumber \\
&&+ \frac{1}{4}\frac{1}{m^2_{\tilde
d_j}}\left[D^d_{2j\ell}D^{d*}_{3j\ell}F_2(x_{\chi^0_\ell \tilde
d_j})+\frac{m_{\chi^0}}{m_b}D^d_{2j\ell}C^{d*}_{3j\ell}F_4(x_{\chi^0_\ell 
\tilde
d_j})\right] \nonumber \\
&&+ \frac{2}{3}g_s^2\frac{1}{m^2_{\tilde
d_j}}\left[\Gamma^{d\dagger}_{L2j}\Gamma^d_{Lj3}F_2(x_{g\tilde
d_j})-\frac{m_g}{m_b}\Gamma^{d\dagger}_{L2j}\Gamma^d_{Rj3}F_4(x_{g\tilde
d_j})\right] \nonumber \\
&&- \frac{3}{4}g_s^2\frac{1}{m^2_{\tilde
d_j}}\left[\Gamma^{d\dagger}_{L2j}\Gamma^d_{Lj3}F_1(x_{g\tilde
d_j})-\frac{m_g}{m_b}\Gamma^{d\dagger}_{L2j}\Gamma^d_{Rj3}F_3(x_{g\tilde
d_j})\right]
\\
\nonumber \\
\tilde C_{7\mathrm{MSSM}} & = &
-\frac{G_{F}}{\sqrt{2}}K_{ts}^{*}K_{tb}\frac{m_sm_b}{m_t^2}\,\tan^2\beta\,
 x_{tH}[Q_{u}F_1(x_{tH})+F_2(x_{tH})] \nonumber \\
 &&+\frac{1}{4}\frac{1}{m^2_{\tilde
 u_j}}A^d_{2j\ell}A^{d*}_{3j\ell}\left[F_1(x_{\chi^{\mathrm{ch}}_\ell \tilde
u_j})+Q_{u}F_2(x_{\chi^{\mathrm{ch}}_\ell \tilde
 u_j})\right] \nonumber\\
&&+\frac{1}{4}\frac{1}{m^2_{\tilde
u_j}}\frac{m_{\chi^{\mathrm{ch}}_\ell}}{m_b}A^d_{2j\ell}B^{d*}_{3j\ell}\left[F_3(x_{\chi^{\mathrm{ch}}_\ell
 \tilde u_j})+Q_u F_4(x_{\chi^{\mathrm{ch}}_\ell
 \tilde u_j})\right] \nonumber \\
&&+ \frac{Q_d}{4}\frac{1}{m^2_{\tilde
 d_j}}\left[C^d_{2j\ell}C^{d*}_{3j\ell}F_2(x_{\chi^0_\ell \tilde
 d_j})+\frac{m_{\chi^0_\ell}}{m_b}C^d_{2j\ell} 
D^{d*}_{3j\ell}F_4(x_{\chi^0_\ell
 \tilde d_j})\right] \nonumber \\
 &&+ \frac{2}{3}Q_d\,g_s^2\frac{1}{m^2_{\tilde
 d_j}}\left[\Gamma^{d\dagger}_{R2j}\Gamma^d_{Rj3}F_2(x_{g\tilde
 d_j})-\frac{m_g}{m_b}\Gamma^{d\dagger}_{R2j}\Gamma^d_{Lj3}F_4(x_{g\tilde
 d_j})\right]
 \\
 \nonumber \\
 \tilde C_{8\mathrm{MSSM}} & = &
 -\frac{G_{F}}{\sqrt{2}}K_{ts}^{*}K_{tb}\frac{m_sm_b}{m_t^2}\,\tan^2\beta\,
x_{tH}F_1(x_{tH}) \nonumber \\
&&+ \frac{1}{4}\frac{1}{m^2_{\tilde
u_j}}\left[A^d_{2j\ell}A^{d*}_{3j\ell}F_2(x_{\chi^{\mathrm{ch}}_\ell \tilde
u_j})+\frac{m_{\chi^{\mathrm{ch}}_\ell}}{m_b}A^d_{2j\ell}B^{d*}_{3j\ell}F_4(x_{\chi^{\mathrm{ch}}_\ell
\tilde u_j})\right] \nonumber \\
&&+ \frac{1}{4}\frac{1}{m^2_{\tilde
d_j}}\left[C^d_{2j\ell}C^{d*}_{3j\ell}F_2(x_{\chi^0_\ell \tilde
d_j})+\frac{m_{\chi^0_\ell}}{m_b}C^d_{2j\ell}D^{d*}_{3j\ell}F_4(x_{\chi^0_\ell 
\tilde
d_j})\right] \nonumber \\
&&+ \frac{2}{3}g_s^2\frac{1}{m^2_{\tilde
d_j}}\left[\Gamma^{d\dagger}_{R2j}\Gamma^d_{Rj3}F_2(x_{g\tilde
d_j})-\frac{m_g}{m_b}\Gamma^{d\dagger}_{R2j}\Gamma^d_{Lj3}F_4(x_{g\tilde
d_j})\right] \nonumber \\
&&- \frac{3}{4}g_s^2\frac{1}{m^2_{\tilde
d_j}}\left[\Gamma^{d\dagger}_{R2j}\Gamma^d_{Rj3}F_1(x_{g\tilde
d_j})-\frac{m_g}{m_b}\Gamma^{d\dagger}_{R2j}\Gamma^d_{Lj3}F_3(x_{g\tilde
d_j})\right].
 \end{eqnarray}

\subsection{Including the R-parity breaking terms}
\subsubsection{The new operator basis}
As mentioned before, the difference between the standard model and the 
MSSM does
not lie in a change of the operator basis but rather in different 
matching
conditions at $M_W$. This situation changes drastically if one 
includes the
R-parity breaking sector. Now our basis has to be enlarged. To find 
out
which are the relevant new operators we first write down the R-parity 
breaking
Yukawa couplings:

\beq
{\mathcal L}_{\mathrm{Yukawa}\not
R}={\mathcal L}_{\lambda}+{\mathcal L}_{\lambda'}+{\mathcal L}_{\lambda''},
\eeq
where

\begin{eqnarray}
{\mathcal L}_{\lambda} & = & \frac{1}{2}\lambda_{ijk} 
\left[\Gamma^{e\dagger}_{Li\ell}\tilde
e_\ell \overline{e_R}_k\nu_{Lj}+\Gamma^{\nu\dagger}_{j\ell}\tilde 
\nu_\ell
\overline{e_R}_k e_{Li}+\Gamma^e_{R\ell k}\tilde e^\dagger_\ell 
\overline
{e^c_R}_i\nu_{Lj}\right] \nonumber \\
&&+ \frac{1}{2}\lambda^*_{ijk}\left[\Gamma^e_{L\ell i}\tilde 
e^\dagger_\ell
\overline{\nu_L}_j e_{Rk}+\Gamma^{\nu\dagger}_{\ell 
j}\tilde\nu^\dagger_\ell
\overline{e_L}_i e_{Rk}+\Gamma^{e\dagger}_{Rk\ell}\tilde
e_\ell\overline{\nu_L}_j e^c_{Ri}\right] \label{L1} \\ 
\nonumber \\
{\mathcal L}_{\lambda'} & = &
\frac{1}{2}\lambda'_{ijk}\left[\Gamma^{e\dagger}_{Li\ell}K^\dagger_{jm}\tilde
e_\ell\overline{d_R}_k u_{Lm}+\Gamma^{u\dagger}_{Lj\ell}\tilde
u_{\ell}\overline{d_R}_k e_{Li}+\Gamma^d_{R\ell k}K^\dagger_{jm}\tilde
d^\dagger_\ell \overline{e^c_R}_iu_{Lm} \right.\nonumber \\
&&-\left.\Gamma^{\nu\dagger}_{i\ell}\tilde\nu_\ell\overline{d_R}_k
d_{Lj}-\Gamma^{d\dagger}_{Lj\ell}\tilde d_\ell
\overline{d_R}_k\nu_{Li}-\Gamma^d_{R\ell k}\tilde d^\dagger_\ell
\overline{\nu^c_R}_i d_{Lj}\right] \nonumber \\ &&+
\frac{1}{2}\lambda'^*_{ijk}\left[\Gamma^e_{L\ell i}K_{mj}\tilde
e^\dagger_\ell
\overline{u_L}_m d_{Rk}+\Gamma^u_{L\ell j}\tilde u^\dagger_\ell 
\overline{e_L}_i
d_{Rk}+\Gamma^{d\dagger}_{Rk\ell}K_{mj}\tilde d_\ell \overline{u_L}_m 
e^c_{Ri}
\right.\nonumber \\
&&-\left.\Gamma^\nu_{\ell i}\tilde\nu^\dagger_\ell \overline{d_L}_j
d_{Rk}-\Gamma^d_{L\ell j}\tilde d^\dagger_\ell \overline{\nu_L}_i
d_{Rk}-\Gamma^{d\dagger}_{Rk\ell}\tilde d_\ell \overline{d_L}_j
\nu^c_{Ri}\right] \label{L2} \\
\nonumber \\
{\mathcal L}_{\lambda''} & = &
-\frac{1}{4}\lambda''_{ijk}\left[\Gamma^u_{R\ell i}\tilde 
u^\dagger_\ell
\overline{d_R}_j d^c_{Lk}+\Gamma^d_{R\ell j}\tilde d^\dagger_\ell
\overline{u_R}_i d^c_{Lk}+\Gamma^d_{R\ell k}\tilde d_\ell^\dagger
\overline{u_R}_i d^c_{Lj}\right] \nonumber \\
&&- 
\frac{1}{4}\lambda''^*_{ijk}\left[\Gamma^{u\dagger}_{Ri\ell}\tilde
u_\ell
\overline{d^c_L}_k
d_{Rj}+\Gamma^{d\dagger}_{Rj\ell}\tilde d_\ell\overline{d^c_L}_k
u_{Ri}+\Gamma^{d\dagger}_{Rk\ell}\tilde d_\ell\overline{d^c_L}_j 
u_{Ri}\right].
\label{L3}
\end{eqnarray}

Here, all the fields belong to the mass basis. The colour indices 
have been
omitted.\\

The next task is  to build four-quark operators out of two Yukawa 
couplings
that contribute at $O(\alpha_s)$ to $b\rightarrow s\gamma$. The boson 
serves as
a bridge between the fermions in analogy to the $W$ boson in the 
standard model.
The following points must be taken care of:
\begin{itemize}
\item
The top quark is not at our disposal in the five flavour effective 
theory. 
\item
We not only need an ingoing $b$ and an outgoing $s$. The two 
remaining quarks
must be of the same type because one has to be able to close the loop 
with these
fermions.
\end{itemize}

\begin{figure}[bp]

\begin{center}
\epsfbox{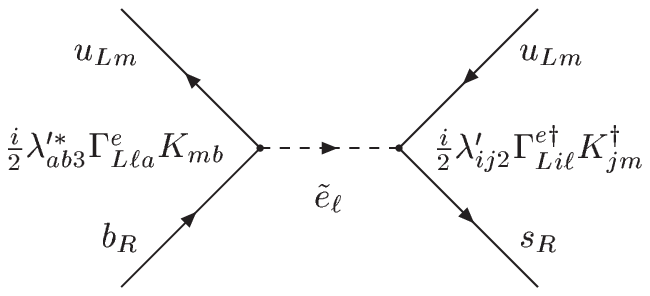}
\end{center}
\caption{\label{fig:7} Illustration of the example in equation 
(\ref{expl}).}
\end{figure}

It is clear that ${\mathcal L}_\lambda$ cannot  participate because it 
contains
no squarks and semi-leptonic operators can be neglected. Also,
${\mathcal L}_{\lambda'}$ does not mix with ${\mathcal L}_{\lambda''}$. 
As an
example, we take the first term of ${\mathcal L}_{\lambda'}$ together 
with his
hermitian conjugate. The situation is depicted in Fig.\ \ref{fig:7}:
\begin{eqnarray}
\lefteqn{\frac{i}{2}\lambda'_{ij2}\Gamma^{e\dagger}_{Li\ell}K^\dagger_{jm}\tilde
e_\ell\overline{s_R} u_{Lm}\frac{i}{k^2-m_{\tilde
e_\ell}^2}\frac{i}{2}\lambda'^*_{ab3}\Gamma^e_{L\ell a}K_{mb}\tilde 
e^\dagger_\ell 
\overline{u_L}_m b_{R}} \nonumber \\ \nonumber \\
&\stackrel{k^2\ll m^2_{\tilde e_\ell}}\longrightarrow&\frac{i}{4}\frac{1}{m_{\tilde
e_\ell}^2}\lambda'_{ij2}\lambda'^*_{aj3}\Gamma^{e\dagger}_{Li\ell}\Gamma^e_{L\ell
a}\;(\overline{s_R}u_{Lj})\;(\overline{u_{L}}_jb_R) \nonumber \\
\nonumber \\
&=&-\frac{i}{8}\frac{1}{m_{\tilde
e_\ell}^2}\lambda'_{ij2}\lambda'^*_{aj3}\Gamma^{e\dagger}_{Li\ell}\Gamma^e_{L\ell
a}\;(\overline{s_R}_\alpha\gamma^\mu
b_{R\beta})\;(\overline{u_L}_{j\alpha}\gamma^\mu u_{Lj\beta})\label{expl}
\end{eqnarray}
In the first step we used the fact that the two squarks have to be of 
the same
type and the unitarity of the CKM-matrix. In the second step we
performed a Fierz rearrangement. This is done to get the same structure 
(i.e., two
vectors) for the four-fermion operator as in the standard model case. The 
advantages
of this rearrangement will become clear when calculating the 
$\gamma$-matrix. In
the last line we put the colour indices $\alpha$, $\beta$ for 
clarity.\\
As one can see clearly, the effect of the (unitary) squark mixing 
matrix
$\Gamma^e_L$ becomes enhanced if the masses of the selectrons are 
very different
for the three generations. If there were a mass degeneracy they would 
simply give
a factor $\delta_{ia}$. This is a general feature in our 
calculations.\\
The operator that appears in Eq.\ (\ref{expl}) is of a new type. It 
consists of a
right-handed $b$- and a right-handed $s$-quark. (Actually, these are 
two operators,
one with a pair of $u$-quarks and the other with two $c$-quarks.)\\

A careful investigation results in the following set of new 
operators:\\
From ${\mathcal L}_\lambda'$ one gets

\begin{eqnarray}
P_1 & = & (\overline{s_R}_\alpha\gamma^\mu
b_{R\beta})\;(\overline{u_L}_\beta\gamma_\mu u_{L\alpha})\nonumber \\
\nonumber \\
P_2 & = & (\overline{s_R}_\alpha\gamma^\mu
b_{R\beta})\;(\overline{c_L}_\beta\gamma_\mu c_{L\alpha}) \nonumber \\
\nonumber \\
P_3 & = & (\overline{s_R}_\alpha\gamma^\mu
b_{R\beta})\;(\overline{d_L}_\beta\gamma_\mu d_{L\alpha}) \nonumber \\
\nonumber \\
P_4 & = & (\overline{s_R}_\alpha\gamma^\mu
b_{R\beta})\;(\overline{s_L}_\beta\gamma_\mu s_{L\alpha}) \nonumber \\
\nonumber \\
P_5 & = & (\overline{s_R}_\alpha\gamma^\mu
b_{R\beta})\;(\overline{b_L}_\beta\gamma_\mu b_{L\alpha}) \nonumber \\
\nonumber \\
P_6 & = & (\overline{s_L}_\alpha\gamma^\mu
b_{L\beta})\;(\overline{d_R}_\beta\gamma_\mu d_{R\alpha}) \nonumber \\
\nonumber \\
P_7 & = & (\overline{s_L}_\alpha\gamma^\mu
b_{L\beta})\;(\overline{s_R}_\beta\gamma_\mu s_{R\alpha}) \nonumber \\
\nonumber \\
P_8 & = & (\overline{s_L}_\alpha\gamma^\mu
b_{L\beta})\;(\overline{b_R}_\beta\gamma_\mu b_{R\alpha}) \nonumber \\
\nonumber \\
P_9 & = & (\overline{s_R}_\alpha\gamma^\mu
b_{R\alpha})\;\sum_i(\overline{q_R}_{i\beta}\gamma^\mu q_{Ri\beta}) \nonumber \\
\nonumber \\
P_{10} & = & (\overline{s_R}_\alpha\gamma^\mu
b_{R\beta})\;\sum_i(\overline{q_R}_{i\beta}\gamma^\mu q_{Ri\alpha}) \nonumber \\
\nonumber \\
P_{11} & = & (\overline{s_R}_\alpha\gamma^\mu
b_{R\alpha})\;\sum_i(\overline{q_L}_{i\beta}\gamma^\mu q_{Li\beta}) \nonumber \\
\nonumber \\
P_{12} & = & (\overline{s_R}_\alpha\gamma^\mu
b_{R\beta})\;\sum_i(\overline{q_L}_{i\beta}\gamma^\mu q_{Li\alpha}).
\end{eqnarray}
It is worth noting that the operators $P_1$ - $P_8$ emerge directly 
from the
Lagrangian and $P_9$ - $P_{12}$ are induced through QCD corrections. 
One would
expect partners of $P_1$ - $P_8$ with colour structure
$(\alpha\alpha)(\beta\beta)$ to be introduced by QCD. This does
not happen at leading-log (accidently).\\

${\mathcal L}_{\lambda''}$ leads to the following additional operators:

\begin{eqnarray}
R_1 & = & (\overline{s_R}_\alpha\gamma^\mu
b_{R\alpha})\;(\overline{u_R}_\beta\gamma_\mu u_{R\beta}) \nonumber \\
\nonumber \\
R_2 & = & (\overline{s_R}_\alpha\gamma^\mu
b_{R\beta})\;(\overline{u_R}_\beta\gamma_\mu u_{R\alpha}) \nonumber \\
\nonumber \\
R_3 & = & (\overline{s_R}_\alpha\gamma^\mu
b_{R\alpha})\;(\overline{c_R}_\beta\gamma_\mu c_{R\beta}) \nonumber \\
\nonumber \\
R_4 & = & (\overline{s_R}_\alpha\gamma^\mu
b_{R\beta})\;(\overline{c_R}_\beta\gamma_\mu c_{R\alpha}) \nonumber \\
\nonumber \\
R_5 & = & (\overline{s_R}_\alpha\gamma^\mu
b_{R\alpha})\;(\overline{d_R}_\beta\gamma_\mu d_{R\beta}) \nonumber \\
\nonumber \\
R_6 & = & (\overline{s_R}_\alpha\gamma^\mu
b_{R\beta})\;(\overline{d_R}_\beta\gamma_\mu d_{R\alpha}).
\end{eqnarray}
Here, all the operators $R_1$ - $R_6$ appear already in the 
tree-level effective
Lagrangian.\\
In principle, the effective Hamiltonian also contains semi-leptonic Operators.
However, these can be neglected because, as a consequence of their Dirac
structure, they do not contribute to the decay $b\rightarrow s\gamma$.

\subsubsection{The $\gamma$-matrix}\label{gamma}
\begin{figure}[tbp]

\begin{center}
\epsfbox{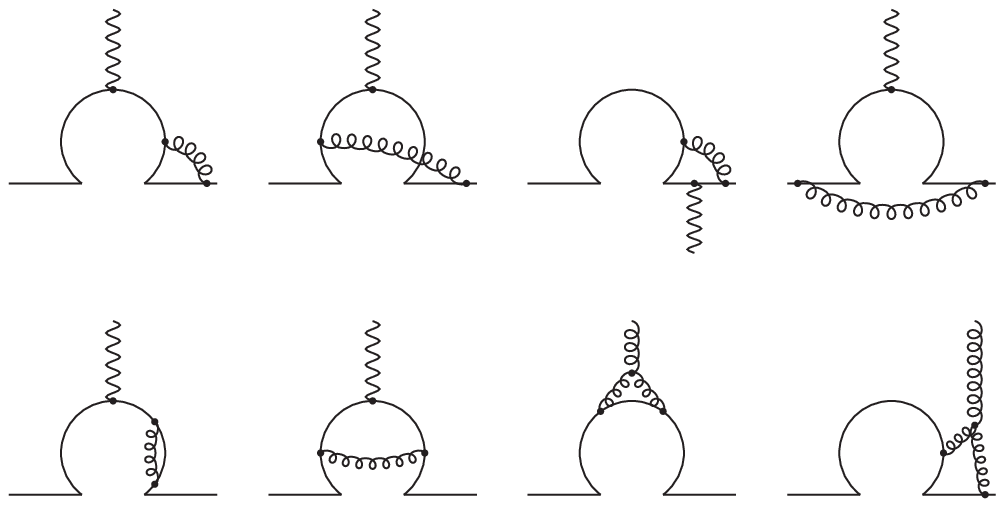}
\end{center}
\caption{\label{fig:4} Two-loop diagrams needed for calculating the
$\gamma$-matrix. The wavy line can be a photon or a gluon. These 
diagrams have
\emph{no closed fermion loop}.}
\end{figure}

\begin{figure}[tp]

\begin{center}
\epsfbox{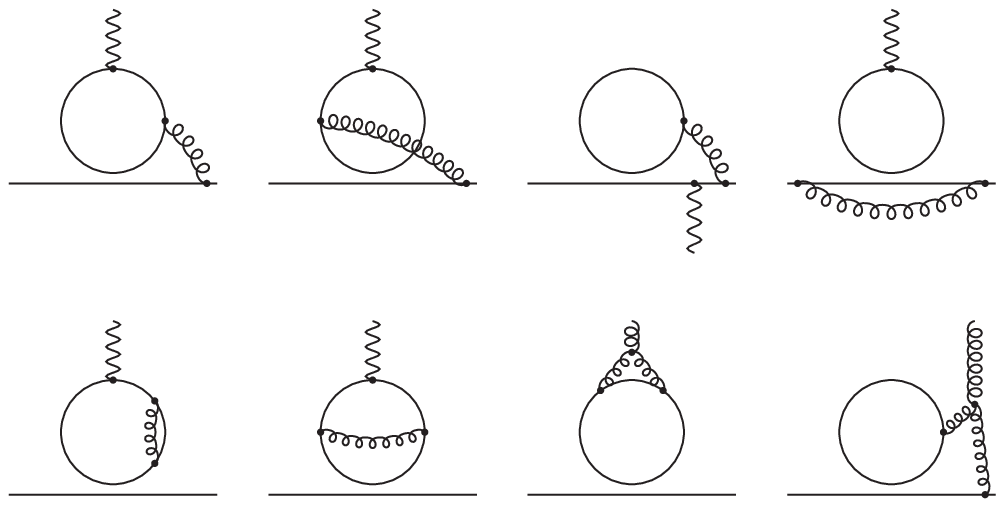}
\end{center}
\caption{\label{fig:5} Two-loop diagrams needed for calculating the
$\gamma$-matrix. The wavy line can be a photon or a gluon. These 
diagrams
\emph{have} a closed fermion loop.}
\end{figure}

The whole basis now consists of 28 operators. Their QCD-mixing is 
described by a
$28\times 28$-$\gamma$-matrix. It is depicted in appendix 
\ref{agamma}. There are three
different blocks in this matrix that have to be treated in separate 
ways.
\begin{itemize}
\item
\emph{mixing of four-fermion operators among themselves}\\
This block involves the one-loop diagrams of Fig.\ \ref{fig:2}. There 
are no
further complications. The mixing of $O_1$ - $O_8$ is known already 
since a long time \cite{gilman}.

\item
\emph{mixing of $O_7$, $O_8$ and $\tilde O_7$, $\tilde O_8$ among 
themselves}\\
These entries need not to be computed. The mixing of $O_7$ and $O_8$ 
is known
and the new operators mix in exactly the same way such that the 
corresponding
numbers can be copied. 

\item
\emph{mixing of the four-fermion operators with $O_7$, $O_8$, $\tilde 
O_7$,
$\tilde O_8$}\\
This task is more difficult. In principle, one has to compute the 
divergent part
of all the diagrams of Fig.\ \ref{fig:4} and \ref{fig:5} together 
with their counterterms and the
contributions of the evanescent operators \cite{chetyrkin}. Moreover, there are four 
different
types of chiralities to be inserted: $(LL)(LL)$, $(LL)(RR)$, 
$(RR)(LL)$ and
$(RR)(RR)$. For the first two this calculation had to be performed 
for the case
of the standard model. The detailed results are listed in \cite{ciuchini1}. 
The second
two types of insertions are new. Fortunately, one can deduce the 
divergent parts of
these types of insertions by making the following two observations: 
\begin{itemize}
\item
The diagrams of Fig.\ \ref{fig:4} contain only one fermion line. 
Here, it
is crucial if the two quark pairs of the inserted operator have the 
same or
opposite chirality. Thus, an insertion of a $(RR)(RR)$ leads to the 
same
divergence as an insertion of a $(LL)(LL)$ as well as the divergence 
of an
insertion of a $(RR)(LL)$ is the same as for the case of $(LL)(RR)$.
\item
One must be careful with the diagrams that contain a closed fermion loop 
(Fig.\ \ref{fig:5}). At a first sight, one may deduce that the
divergence does depend on the chirality of the
quarks running in the loop but \emph{not} on the chirality of the 
$s$- and the $b$-quark. 
However, this is wrong.\footnote{We thank the authors of \cite{chun} for
clarifying this point to us.} The difference between a right- and a left-handed
quark in the loop results in a term
$\mathrm{Tr}(\pm\gamma^5\,\Gamma)$ where $\Gamma$ stands for a collection of at
least four Dirac $\gamma$-matrices and the sign corresponds to a right- or left-handed quark
in the loop, respectively. The trace leads to an
$\varepsilon$-tensor that is contracted with a $\gamma$-matrix between the external
quarks. This produces an additional $\gamma^5$. Now note that $\gamma^5\, P_R=+P_R$,
whereas $\gamma^5\,P_L=-P_L$. This means, we need to change the
chirality in \emph{both} fermion pairs to end up with the same mixing. Hence, if
an insertion of an operator of type $(LL)(LL)$ [$(LL)(RR)$] gives a certain
contribution to $O_{7/8}$ the  $(RR)(RR)$ [$(RR)(LL)$] operator will give exactly 
the same contribution  to $\tilde O_{7/8}$.
\end{itemize}
\end{itemize}
To summarize, the mixing of the four-fermion operators with $\tilde 
O_{7/8}$ can
be deduced completely from the results of \cite{ciuchini1} by interchanging
left-handed and right-handed projectors.\\

All the previous calculations involve $\gamma^5$. It is therefore 
clear that the
results depend on the regularization scheme. This dependence will be 
cancelled by
finite one-loop (but $O(\alpha_s^0)$) contributions of some 
four-fermion
operators $Q_i$ to the Amplitude $A$ of $b\rightarrow s\gamma$ 
through the
diagram of Fig.\ \ref{fig:3}. Schematically, the result is
then
\beq
A=C_7\langle s\gamma |O_7|b\rangle_{\mathrm{tree}}+\tilde C_7\langle s\gamma|\tilde
O_7|b\rangle_{\mathrm{tree}}+\sum_i C_{Q_i}\langle s\gamma |Q_i|b\rangle_{\mathrm{1-loop}}.
\eeq
An alternative \cite{buras3} is to define effective coefficients in a way that $A$ 
becomes
\beq
A=C_7^{\mathrm{eff}}\langle s\gamma |O_7|b\rangle_{\mathrm{tree}}+\tilde
C_7^{\mathrm{eff}}\langle s\gamma|\tilde O_7|b\rangle_{\mathrm{tree}}.
\eeq
This can be achieved by defining four vectors $\{y_i\}$, $\{z_i\}$, 
$\{\tilde
y_i\}$,
and $\{\tilde z_i\}$ through 
\begin{eqnarray}
\langle s\gamma|Q_i|b\rangle _{\mathrm{1-loop}}
&=:&y_i\langle s\gamma|O_7|b\rangle _{\mathrm{tree}}
\nonumber \\ \nonumber \\
\langle s\:\mathrm{gluon}|Q_i|b\rangle _{\mathrm{1-loop}}
&=:&z_i\langle s\:\mathrm{gluon}|O_8|b\rangle _{\mathrm{tree}}
\nonumber \\ \nonumber \\
\langle s\gamma|Q_i|b\rangle _{\mathrm{1-loop}}&=:&\tilde y_i\langle s\gamma|\tilde 
O_7|b\rangle _{\mathrm{tree}}\nonumber \\ \nonumber \\
\langle s\:\mathrm{gluon}|Q_i|b\rangle _{\mathrm{1-loop}}&=:&\tilde 
z_i\langle s\:\mathrm{gluon}|\tilde
O_8|b\rangle _{\mathrm{tree}}.
\end{eqnarray}
The effective Wilson-coefficients must be defined as

\begin{eqnarray}
C_7^{\mathrm{eff}}(\mu)&:=&C_7(\mu)+\sum_i y_i C_i(\mu)\nonumber \\
\nonumber \\
C_8^{\mathrm{eff}}(\mu)&:=&C_8(\mu)+\sum_i z_i C_i(\mu) \nonumber \\
 \nonumber \\
\tilde C_7^{\mathrm{eff}}(\mu)&:=&\tilde C_7(\mu)+\sum_i\tilde y_i 
C_i(\mu)\nonumber \\
\nonumber \\
\tilde C _8^{\mathrm{eff}}(\mu)&:=&\tilde C_8(\mu)+\sum_i\tilde z_i 
C_i(\mu)\nonumber \\
\nonumber \\
C_{Q_i}^{\mathrm{eff}}(\mu)&:=&C_{Q_i}(\mu).
\end{eqnarray}
The vector 
\beq
\vec
C^{\mathrm{eff}}(\mu):=\left\{C_{Q_i}^{\mathrm{eff}}(\mu),C_7^{\mathrm{eff}}(\mu),C_8^{\mathrm{eff}}(\mu),
\tilde C_7^{\mathrm{eff}}(\mu),\tilde C_8^{\mathrm{eff}}(\mu)\right\}
\eeq
is then regularization scheme independent. Remember that the index 
$i$ runs over
all four fermion operators.\newline
The effective coefficients obey RGE's
which can be derived from the RGE's for $C_k(\mu)$, where $k$ labels 
the whole set
of operators. They are
\beq
\frac{d}{\ln\mu}C_k^{\mathrm{eff}}(\mu)=\frac{\alpha_s}{4\pi}\gamma_{jk}^
{\mathrm{eff}}\,C_j^{\mathrm{eff}}(\mu),
\eeq
where
\beq
\gamma _{jk}^{\mathrm{eff}}=\left\{\begin{array}{ll}
\gamma_{jk}+\sum_{i=1}^{24}y_i\gamma_{ji}-y_j\gamma_{O_7O_7}
-z_j\gamma_{O_8O_7}&k=O_7,\:j=1,\ldots,24\\
\gamma_{jk}+\sum_{i=1}^{24}z_i\gamma_{ji}-z_j\gamma_{O_8O_8}
&k=O_8,\:j=1,\ldots,24\\
\gamma_{jk}+\sum_{i=1}^{24}\tilde y_i\gamma_{ji}-\tilde 
y_j\gamma_{\tilde
O_7\tilde O_{7}}-\tilde z_j\gamma_{\tilde O_8\tilde O_7}&k=\tilde
O_7,\:j=1,\ldots,24\\
\gamma_{jk}+\sum_{i=1}^{24}\tilde z_i\gamma_{ji}-\tilde 
z_j\gamma_{\tilde
O_8\tilde O_8}&k=\tilde O_8,\:j=1,\ldots,24\\
\gamma_{jk}&\mbox{otherwise}
\end{array}\right.
\eeq
For a finite contribution of an operator inserted in the diagram of 
Fig.\ \ref{fig:3} we need
\begin{itemize}
\item
two pairs of fermions with different chirality, i.e., operators of the
form $(LL)(RR)$ or $(RR)(LL)$
\item
a $b$-quark running in the loop.
\end{itemize}
This reduces  the possibilities to $O_5$, $O_6$, $P_5$, $P_8$, 
$P_{11}$ and
$P_{12}$. The results for $\{y_i\}$, $\{z_i\}$, $\{\tilde
y_i\}$ and $\{\tilde z_i\}$ are then
\beq
\begin{array}{lcllcl}
y_i&=&\left\{\begin{array}{cl}-\frac{2}{3}&i=O_5\\
				-2&i=O_6,\:P_8\\
				0&\mbox{otherwise}
	\end{array}\right.
&z_i&=&\left\{\begin{array}{cl}2&i=O_5\\
				0&\mbox{otherwise}
	\end{array}\right.\\
	\\
\tilde y_i&=&\left\{\begin{array}{cl}-\frac{2}{3}&i=P_{11}\\
				-2&i=P_5,\:P_{12}\\
				0&\mbox{otherwise}
	\end{array}\right.
&\tilde z_i&=&\left\{\begin{array}{cl}2&i=P_{11}\\
				0&\mbox{otherwise}
	\end{array}\right.
\end{array}
\eeq

There is one more subtlety concerning the matching at $M_W$: In the standard
model we have $\langle s\gamma|H^{\mathrm{eff}}|b\rangle =C_7(M_W)\langle
s\gamma|O_7|b\rangle$. Now, more operators contribute:
\begin{eqnarray}
\langle s\gamma|H^{\mathrm{eff}}|b\rangle&=&C_7(M_W)\langle s\gamma|O_7|b\rangle
+\tilde C_7(M_W)\langle s\gamma|\tilde O_7|b\rangle \nonumber \\
&&-2C_{P_8}(M_W)\langle s\gamma|P_8|b\rangle -2C_{P_5}(M_W)\langle s\gamma|P_5|b\rangle
\nonumber \\
&=&C_7^{\mathrm{eff}}(M_W)\langle s\gamma|O_7|b\rangle +\tilde
C_7^{\mathrm{eff}}(M_W)\langle s\gamma|\tilde O_7|b\rangle.
\end{eqnarray}
Hence, the matching procedure does not lead to $C_7(M_W)$ and $\tilde C_7(M_W)$
but directly to $C_7^{\mathrm{eff}}(M_W)$ and $\tilde C_7^{\mathrm{eff}}(M_W)$.

\subsubsection{The matching conditions}

The matching for the additional four-fermion operators must only be 
performed at
tree-level. The matching conditions are therefore easily derived. An 
example is
given in (\ref{expl}). The complete set is

\begin{eqnarray}
C_{P_1}^{\mathrm{eff}}(M_W)&=&-\frac{1}{8}\lambda'_{i12}\lambda'^*_{j13}\Gamma^{e\dagger}_{Li\ell}\Gamma^e_{L\ell
j}\frac{1}{m_{\tilde e_\ell}^2} \nonumber \\
\nonumber \\
C_{P_2}^{\mathrm{eff}}(M_W)&=&-\frac{1}{8}\lambda'_{i22}\lambda'^*_{j23}\Gamma^{e\dagger}_{Li\ell}\Gamma^e_{L\ell
j}\frac{1}{m_{\tilde 
e_\ell}^2}+\frac{G_F}{\sqrt{2}}K_{ts}^*K_{tb}\frac{m_s 
m_b}{2m_{H^+}^2}\tan ^2\beta \nonumber \\
\nonumber \\
C_{P_3}^{\mathrm{eff}}(M_W)&=&-\frac{1}{8}\lambda'_{i12}\lambda'^*_{j13}\Gamma^{\nu\dagger}_{Li\ell}\Gamma^\nu_{L\ell
j}\frac{1}{m_{\tilde \nu_\ell}^2} \nonumber \\
\nonumber \\
C_{P_4}^{\mathrm{eff}}(M_W)&=&-\frac{1}{8}\lambda'_{i22}\lambda'^*_{j23}\Gamma^{\nu\dagger}_{Li\ell}\Gamma^\nu_{L\ell
j}\frac{1}{m_{\tilde \nu_\ell}^2} \nonumber \\
\nonumber \\
C_{P_5}^{\mathrm{eff}}(M_W)&=&-\frac{1}{8}\lambda'_{i32}\lambda'^*_{j33}\Gamma^{\nu\dagger}_{Li\ell}\Gamma^\nu_{L\ell
j}\frac{1}{m_{\tilde \nu_\ell}^2} \nonumber \\
\nonumber \\
C_{P_6}^{\mathrm{eff}}(M_W)&=&-\frac{1}{8}\lambda'_{i31}\lambda'^*_{j21}\Gamma^{\nu\dagger}_{Li\ell}\Gamma^\nu_{L\ell
j}\frac{1}{m_{\tilde \nu_\ell}^2} \nonumber \\
\nonumber \\
C_{P_7}^{\mathrm{eff}}(M_W)&=&-\frac{1}{8}\lambda'_{i32}\lambda'^*_{j22}\Gamma^{\nu\dagger}_{Li\ell}\Gamma^\nu_{L\ell
j}\frac{1}{m_{\tilde \nu_\ell}^2} \nonumber \\
\nonumber \\
C_{P_8}^{\mathrm{eff}}(M_W)&=&-\frac{1}{8}\lambda'_{i33}\lambda'^*_{j23}\Gamma^{\nu\dagger}_{Li\ell}\Gamma^\nu_{L\ell
j}\frac{1}{m_{\tilde \nu_\ell}^2} \nonumber \\
\nonumber \\
C_{P_9}^{\mathrm{eff}}(M_W)&=&C_{P_{10}}^{\mathrm{eff}}(M_W)
=C_{P_{11}}^{\mathrm{eff}}(M_W)=C_{P_{12}}^{\mathrm{eff}}(M_W)=0 \nonumber \\
\nonumber \\
\nonumber \\
C_{R_1}^{\mathrm{eff}}(M_W)&=&\frac{1}{8}\lambda''_{1i2}\lambda''^*_{1j3}\Gamma^{d\dagger}_{Rj\ell}\Gamma^d_{R\ell
i}\frac{1}{m_{\tilde d_\ell}^2} \nonumber \\
\nonumber \\
C_{R_2}^{\mathrm{eff}}(M_W)&=&-\frac{1}{8}\lambda''_{1i2}\lambda''^*_{1j3}\Gamma^{d\dagger}_{Rj\ell}\Gamma^d_{R\ell
i}\frac{1}{m_{\tilde d_\ell}^2}=-C_{R_1}^{\mathrm{eff}}(M_W) \nonumber \\
\nonumber \\
C_{R_3}^{\mathrm{eff}}(M_W)&=&\frac{1}{8}\lambda''_{2i2}\lambda''^*_{2j3}\Gamma^{d\dagger}_{Rj\ell}\Gamma^d_{R\ell
i}\frac{1}{m_{\tilde d_\ell}^2} \nonumber \\
\nonumber \\
C_{R_4}^{\mathrm{eff}}(M_W)&=&-\frac{1}{8}\lambda''_{2i2}\lambda''^*_{2j3}\Gamma^{d\dagger}_{Rj\ell}\Gamma^d_{R\ell
i}\frac{1}{m_{\tilde d_\ell}^2}=-C_{R_3}^{\mathrm{eff}}(M_W) \nonumber \\
\nonumber \\
C_{R_5}^{\mathrm{eff}}(M_W)&=&\frac{1}{8}\lambda''_{i12}\lambda''^*_{j13}\Gamma^{u\dagger}_{Rj\ell}\Gamma^u_{R\ell
i}\frac{1}{m_{\tilde u_\ell}^2} \nonumber \\
\nonumber \\
C_{R_6}^{\mathrm{eff}}(M_W)&=&-\frac{1}{8}\lambda''_{i12}\lambda''^*_{j13}\Gamma^{u\dagger}_{Rj\ell}\Gamma^u_{R\ell
i}\frac{1}{m_{\tilde u_\ell}^2}=-C_{R_5}^{\mathrm{eff}}(M_W)\label{match}.
\end{eqnarray}
In $C_{P_2}^{\mathrm{eff}}(M_W)$ we included a term coming from Higgs 
exchange because it can possibly be large in the case of a large
$\tan\beta$.\\
There are some more terms to add to $C_7$, $C_8$, $\tilde C_7$, and 
$\tilde C_8$, too
\begin{eqnarray}
C_7^{\mathrm{eff}}(M_W)&=&C_{7\mathrm{MSSM}}(M_W) \nonumber \\
&&-\frac{1}{4}\frac{Q_d}{m_{\tilde\nu_\ell}^2}\lambda'_{i3k}\lambda'^*_{j2k}\Gamma^{\nu\dagger}_{i\ell}\Gamma^\nu_{\ell
j}F_1(x_{d_k\tilde\nu_\ell}) \nonumber \\
&&+\frac{1}{48}\frac{Q_d}{m^2_{\tilde
d_\ell}}\lambda'_{k3i}\lambda'^*_{k2j}\Gamma^{d\dagger}_{Rj\ell}\Gamma^d_{R\ell
i}\\[1cm]
C_8^{\mathrm{eff}}(M_W)&=&C_{8\mathrm{MSSM}}(M_W) \nonumber \\
&&-\frac{1}{4}\frac{1}{m_{\tilde\nu_\ell}^2}\lambda'_{i3k}\lambda'^*_{j2k}\Gamma^{\nu\dagger}_{i\ell}\Gamma^\nu_{\ell
j}F_1(x_{d_k\tilde\nu_\ell}) \nonumber \\
&&+\frac{1}{48}\frac{1}{m^2_{\tilde
d_\ell}}\lambda'_{k3i}\lambda'^*_{k2j}\Gamma^{d\dagger}_{Rj\ell}\Gamma^d_{R\ell
i} 
\\[1cm]
\tilde C_7^{\mathrm{eff}}(M_W)&=&\tilde C_{7\mathrm{MSSM}}(M_W) \nonumber \\
&&+\frac{1}{4}\frac{1}{m_{\tilde 
e_\ell}^2}\lambda'_{ia2}\lambda'^*_{jb3}K^\dagger_{an}K_{nb}\Gamma^{e\dagger}_{Li\ell}\Gamma^e_{L\ell
j}\left[-Q_u F_1(x_{u_n\tilde e_\ell})+Q_e F_2(x_{u_n\tilde 
e_\ell})\right] \nonumber \\
&&+\frac{1}{4}\frac{1}{m_{\tilde 
u_\ell}^2}\lambda'_{ki2}\lambda'^*_{kj3}\Gamma^{u\dagger}_{Li\ell}\Gamma^u_{L\ell
j}\left[-Q_e F_1(x_{e_k\tilde u_\ell})+Q_u F_2(x_{e_k\tilde 
u_\ell})\right] \nonumber \\
&&-\frac{Q_d}{4}\frac{1}{m_{\tilde\nu_\ell}^2}\lambda'_{ik2}\lambda'^*_{jk3}\Gamma^{\nu\dagger}_{i\ell}\Gamma^\nu_{\ell
j}F_1(x_{d_k\tilde\nu_\ell}) \nonumber \\
&&+\frac{Q_d}{48}\frac{1}{m_{\tilde 
d_\ell}^2}\lambda'_{ki2}\lambda'^*_{kj3}\Gamma^{d\dagger}_{Li\ell}\Gamma^d_{L\ell
j} \nonumber \\
&&+\frac{1}{2}\frac{1}{m_{\tilde 
u_\ell}^2}\lambda''_{ik2}\lambda''^*_{jk3}\Gamma^{u\dagger}_{Rj\ell}\Gamma^u_{R\ell
i}\left[Q_d F_1(x_{d_k\tilde u_\ell})-Q_u F_2(x_{d_k\tilde 
u_\ell})\right] \nonumber \\
&&+\frac{1}{2}\frac{1}{m_{\tilde 
d_\ell}^2}\lambda''_{ki2}\lambda''^*_{kj3}\Gamma^{d\dagger}_{Rj\ell}\Gamma^d_{R\ell
i}\left[Q_u F_1(x_{u_k\tilde d_\ell})-Q_d F_2(x_{u_k\tilde 
d_\ell})\right]\\[1cm]
\tilde C_8^{\mathrm{eff}}(M_W)&=&\tilde C_{8\mathrm{MSSM}}(M_W) \nonumber \\
&&-\frac{1}{4}\frac{1}{m_{\tilde 
e_\ell}^2}\lambda'_{ia2}\lambda'^*_{jb3}K^\dagger_{an}K_{nb}\Gamma^{e\dagger}_{Li\ell}\Gamma^e_{L\ell
j}F_1(x_{u_n\tilde e_\ell}) \nonumber \\
&&+\frac{1}{4}\frac{1}{m_{\tilde 
u_\ell}^2}\lambda'_{ki2}\lambda'^*_{kj3}\Gamma^{u\dagger}_{Li\ell}\Gamma^u_{L\ell
j}F_2(x_{e_k\tilde u_\ell}) \nonumber \\
&&-\frac{1}{4}\frac{1}{m_{\tilde\nu_\ell}^2}\lambda'_{ik2}\lambda'^*_{jk3}\Gamma^{\nu\dagger}_{i\ell}\Gamma^\nu_{\ell
j}F_1(x_{d_k\tilde\nu_\ell}) \nonumber \\
&&+\frac{1}{48}\frac{1}{m_{\tilde 
d_\ell}^2}\lambda'_{ki2}\lambda'^*_{kj3}\Gamma^{d\dagger}_{Li\ell}\Gamma^d_{L\ell
j} \nonumber \\
&&+\frac{1}{4}\frac{1}{m_{\tilde 
u_\ell}^2}\lambda''_{ik2}\lambda''^*_{jk3}\Gamma^{u\dagger}_{Rj\ell}\Gamma^u_{R\ell
i}\left[-F_1(x_{d_k\tilde u_\ell})+F_2(x_{d_k\tilde u_\ell})\right] \nonumber \\
&&+\frac{1}{4}\frac{1}{m_{\tilde 
d_\ell}^2}\lambda''_{ki2}\lambda''^*_{kj3}\Gamma^{d\dagger}_{Rj\ell}\Gamma^d_{R\ell
i}\left[-F_1(x_{u_k\tilde d_\ell})+F_2(x_{u_k\tilde d_\ell})\right] \nonumber \\
&&-2C_{P_5}(M_W).
\end{eqnarray}

\subsubsection{$\mathbf{b\rightarrow u|c\;e\bar\nu}$}

\begin{figure}[tbp]

\begin{center}
\epsfbox{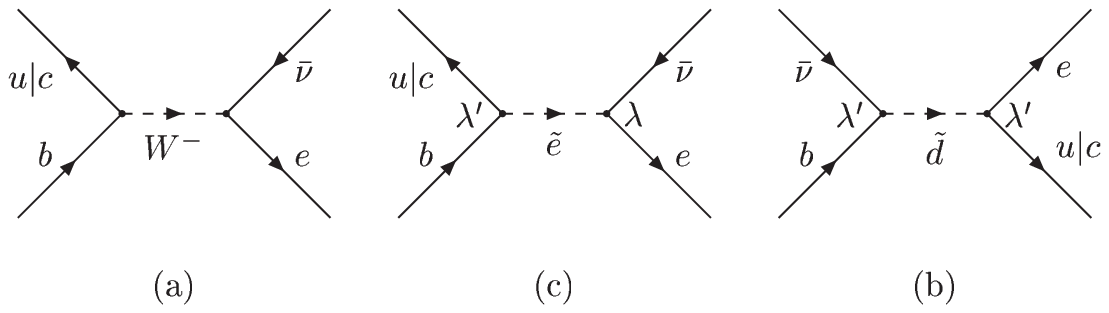}
\end{center}
\caption{\label{fig:8} Contributions to $b\rightarrow u|c\;e\bar\nu$. The arrows
indicate the particle flow.}
\end{figure}

It is convenient to express the branching ratio $\mbox{Br}(b\rightarrow s\gamma)$
through the semi-leptonic decay $b\rightarrow u|c\;e\bar\nu$
\cite{borzumati,cabibbo,cortez,fogli}:
\beq
\mbox{Br}(b\rightarrow s\gamma)=\frac{\Gamma(b\rightarrow s\gamma)}{\Gamma
(b\rightarrow u|c\;e\bar\nu)} \mbox{Br}_{\mathrm{exp}}(b\rightarrow u|c\;e\bar\nu),
\eeq
where we take $\mbox{Br}_{\mathrm{exp}}(b\rightarrow u|c\;e\bar\nu)=10.5\%$
\cite{pdg}.
This has the advantage that the large bottom mass dependence ($\sim m_b^5$) cancels
out. In the SM the semi-leptonic decay is mediated by a $W$-boson [Fig.\ 
\ref{fig:8} (a)] whereas in the case of the MSSM the charged Higgs can take the
role of the $W$. However, the coupling to the leptons is proportional to the
electron mass and hence it can be safely neglected. Introducing the R-parity
breaking terms (\ref{L1}) -(\ref{L3}) offers new decay channels depicted in
Figs.\ \ref{fig:8} (b) and (c) which have to be included in the decay width. Please note
the following few things:
\begin {itemize}
\item
In the MSSM the decay
$b\rightarrow ue\bar \nu$ is suppressed by the small CKM- matrix element
$K_{ub}$ and can therefore be neglected. In our case we have to include
this decay mode.
\item
The absence of lepton generation mixing in the SM forces the anti-neutrino to be
$\bar\nu_e$. This restriction is no longer valid in our case. Therefore, we have 
to sum over the three
generations before squaring the amplitude because
the generation of the neutrino is not detected.  
\item
Setting the mass of the lepton to zero (which is certainly a valid
approximation) the computation in the MSSM does not distinguish between the electron
and the muon. Here, the two particles involve different couplings. We give the
results for an outgoing electron. For the muon just change the ``1'' in the
relevant coupling
to a ``2''.
\end{itemize}
The results are then, to leading order and with $m_u^2/m_b^2=0$

\begin{eqnarray}
\Gamma(b\rightarrow s\gamma)&=&\frac{e^2m_b^5}{192\pi^5}
\frac{3}{4}(|C_7|^2+|\tilde C_7|^2) \\
\Gamma(b\rightarrow u|c\;e\bar\nu)&=&\frac{m_b^5}{192\pi^3}\frac{1}{32}\left\{ \left(
1-8\epsilon^2+8\epsilon^6-\epsilon^8-24\epsilon^4\log\epsilon\right)\times \right.\nonumber \\
&&\left.\left[|2A+C_2|^2+|B_2|^2\right]+|B_1|^2+|C_1|^2\right\},
\end{eqnarray}
where
\begin{eqnarray}
A&=&2\sqrt{2}G_FK_{23}\\
B_r&=&\sum_{i=1}^3\frac{\lambda_{ij1}\lambda'^*_{mn3}}{m^2_{\tilde e_\ell}}
\Gamma^e_{L\ell m}\Gamma^{e\dagger}_{Li\ell}K_{rn}\hspace{1cm} r=1,2\\
C_r&=&\sum_{i=1}^3-\frac{\lambda'_{i3k}\lambda'^*_{1mn}}{m^2_{\tilde d_\ell}}
\Gamma^d_{R\ell k}\Gamma^{d\dagger}_{Rn\ell}K_{rm}\\
\epsilon&=&\frac{m_c}{m_b}.
\end{eqnarray}

\section{Results} \label{results}

The formulas of the previous section are far too complicated to be 
treated ``by
hand'' but it is no problem to feed them to a computer. The 
$\gamma$-matrix is
independent of the parameters of supersymmetry. With the help of
\emph{Mathematica} \cite{mathematica} it is possible to diagonalize it and find the 
influence of
the QCD effects.

\subsection{General results}

Because of the enlarged operator basis the expression for $C_7$ changes. In
general, the solution of the RGE for the Wilson-coefficients is 
given by
\beq
\vec
C^{\mathrm{eff}}(\mu)=V\left[\left(\frac{\alpha_s(M_W)}{\alpha_s(\mu)}\right)^{\frac{\vec
\gamma_D}{2\beta_0}}\right]_DV^{-1}\vec C^{\mathrm{eff}}(M_W),
\eeq
where $V$ diagonalizes $\gamma^T$
\beq
\gamma_D=V^{-1}\gamma^{T} V.
\eeq
$\beta_0=23/3$ is the one-loop beta-function and 
$\vec\gamma_D$ is the vector containing the eigenvalues of $\gamma$. 
In our case
\beq
\vec\gamma_D=\begin{array}[t]{ccccccc}(-16&-16&-16&-16&-16&-16&-16\\
-16&-8&-8&-8&-8&4&4\\
4&4&\frac{28}{3}&\frac{28}{3}&\frac{32}{3}&\frac{32}{3}&2.233\\
2.233&6.266&6.266&-13.791&-13.791&-6.486&-6.486).\end{array}
\eeq

To have an idea which coefficients are relevant we perform a 
numerical {ana-lysis}
with $\alpha_s(M_Z)=0.121$ and $\mu=m_b=4.2$ GeV. The coefficients
$C_7^{\mathrm{eff}}(m_b)$ and $\tilde C_7^{\mathrm{eff}}(m_b)$ are then 
\begin{eqnarray}
C_7^{\mathrm{eff}}(m_b)&=& -0.351 C_{O_2}^{\mathrm{eff}}(M_W)-0.198
C_{P_6}^{\mathrm{eff}}(M_W)-0.198 C_{P_7}^{\mathrm{eff}}(M_W) \nonumber \\
&&-0.178 C_{P_8}^{\mathrm{eff}}(M_W)+0.665 
C_{7}^{\mathrm{eff}}(M_W)+0.093
C_{8}^{\mathrm{eff}}(M_W) \nonumber \\[1cm]
\tilde C_7^{\mathrm{eff}}(m_b)&=&0.510 C_{P_1}^{\mathrm{eff}}(M_W)+0.510
C_{P_2}^{\mathrm{eff}}(M_W)-0.198 C_{P_3}^{\mathrm{eff}}(M_W) \nonumber \\
&&-0.198 C_{P_4}^{\mathrm{eff}}(M_W)-0.178 
C_{P_5}^{\mathrm{eff}}(M_W)+0.381
C_{R_1}^{\mathrm{eff}}(M_W) \nonumber \\
&&+0.381 C_{R_3}^{\mathrm{eff}}(M_W)-0.213 C_{R_5}^{\mathrm{eff}}(M_W)+
0.665 \tilde C_{7}^{\mathrm{eff}}(M_W)\nonumber \\
&&+0.093 \tilde C_{8}^{\mathrm{eff}}(M_W),
\end{eqnarray}
It is clear that the four-fermion operators including a left-handed $s$-quark
contribute to $C_{7/8}$ whereas the ones with a right-handed $s$-quark
contribute to $\tilde C_{7/8}$. The numbers multiplying
the different Wilson coefficients are all of the same size, hence there 
is a priory no term which can be neglected. 

\subsection{Specific results for R-parity violation}

\begin{figure}[tbp]

\begin{center}
\epsfbox{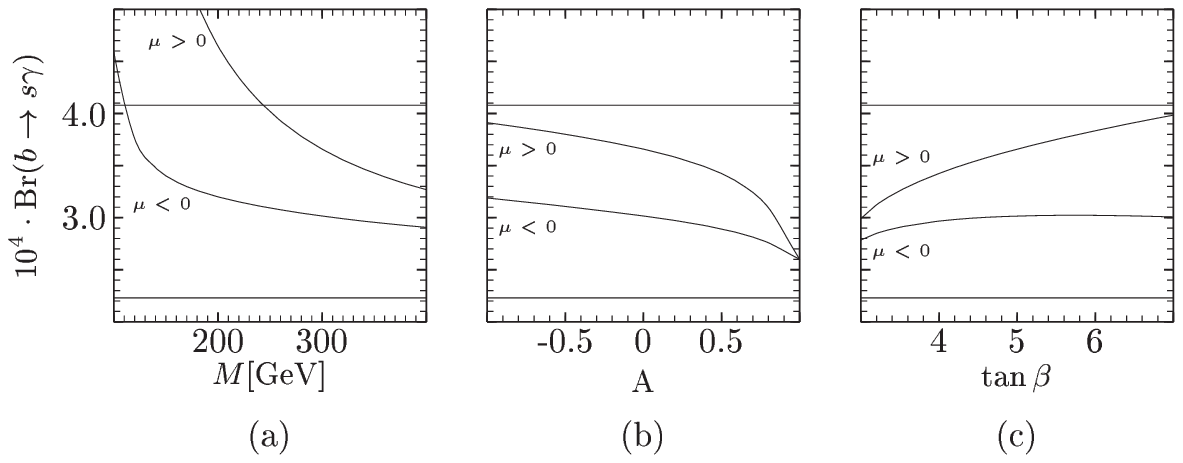}
\end{center}
\caption{\label{fig:9} Behaviour of Br$(b\rightarrow s\gamma)$ in the
neighbourhood of our reference model with $\tan\beta
=5$, $A=0$, $\tilde\mu_{2i}=0$, $M=300$ GeV and, in addition,
$\lambda'_{132}=\lambda'_{122}=0.1$. The horizontal lines show the current experimental 
bounds.}

\end{figure}

\begin{figure}[tbp]

\begin{center}
\epsfbox{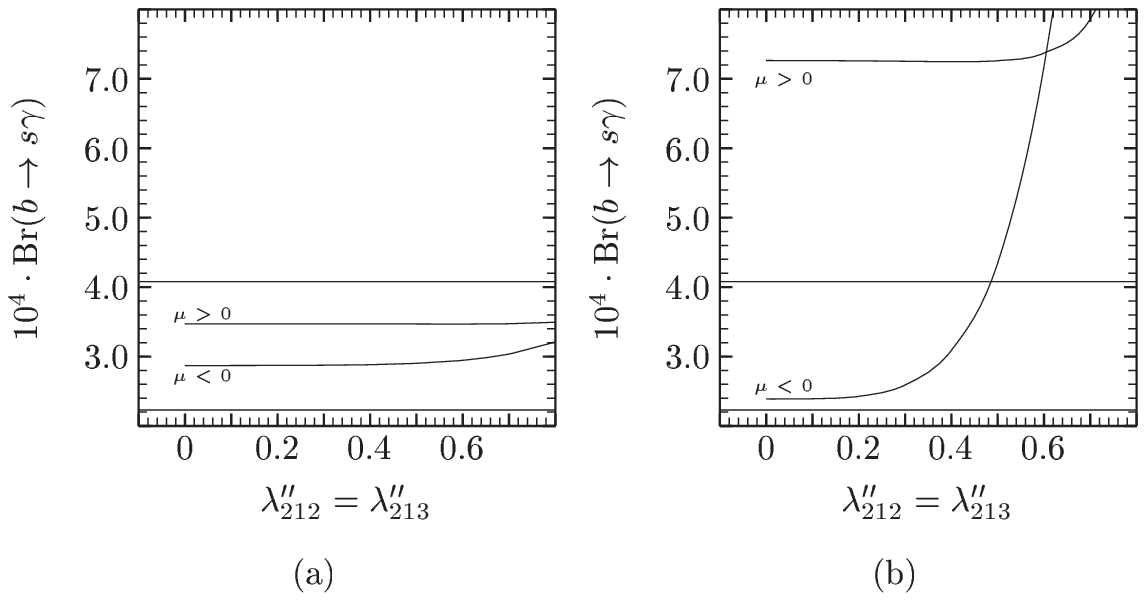}
\end{center}
\caption{\label{fig:10}Values for Br$(b\rightarrow s\gamma)$ as a function of
$\lambda''_{212}=\lambda''_{213}$. The other parameters are: $\tan\beta
=5$, $A=0$, $\tilde\mu_{2i}=0$, and (a): $M=300$ GeV, (b): $M=100$ GeV. The
current bounds are:
$\lambda''_{212}\leq1.23$, $\lambda''_{213}\leq1.23$ \protect\cite{bounds1},
$|\lambda''_{212}\cdot\lambda''_{213}|\leq0.006$ \protect\cite{bounds2}.}

\end{figure}

\begin{figure}[tbp]

\begin{center}
\epsfbox{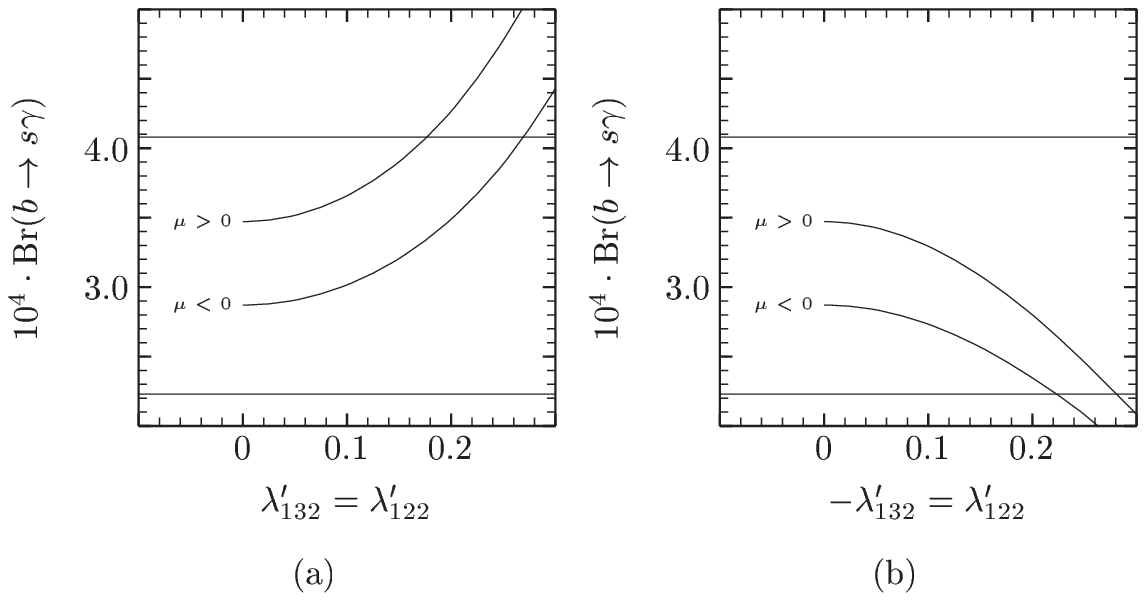}
\end{center}
\caption{\label{fig:12} Values for Br$(b\rightarrow s\gamma)$ as a function of
(a) $\lambda'_{132}=\lambda'_{122}$, (b) $-\lambda'_{132}=\lambda'_{122}$. 
The current bounds are: $\lambda'_{132}\leq0.28 \frac{m_{\tilde{t}}}{100 \mathrm{GeV}}$,
$\lambda'_{122}\leq0.049 \frac{m_{\tilde{d}}}{100 \mathrm{GeV}}$ \protect\cite{bounds1}.}

\end{figure}

\begin{figure}[tbp]

\begin{center}
\epsfbox{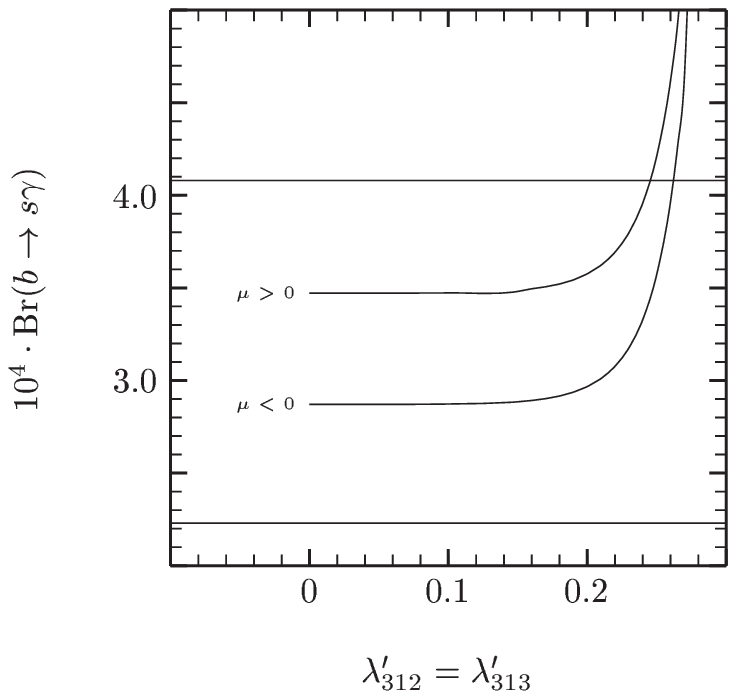}
\end{center}
\caption{\label{fig:11} Values for Br$(b\rightarrow s\gamma)$ as a function of
$\lambda'_{312}=\lambda'_{313}$ within our reference model. The current bounds
are:
$\lambda'_{312}\leq0.11 \frac{m_{\tilde{s}}}{100 \mathrm{GeV}}$, 
$\lambda'_{313}\leq0.11 \frac{m_{\tilde{b}}}{100 \mathrm{GeV}}$ \protect\cite{bounds1}, 
$|\lambda'_{312}\cdot\lambda'_{313}|\leq0.01$ for squarks of 100 GeV
\protect\cite{bounds3}.
 The peak results from a small lightest selectron mass.}
\end{figure}

\begin{figure}[tbp]

\begin{center}
\epsfbox{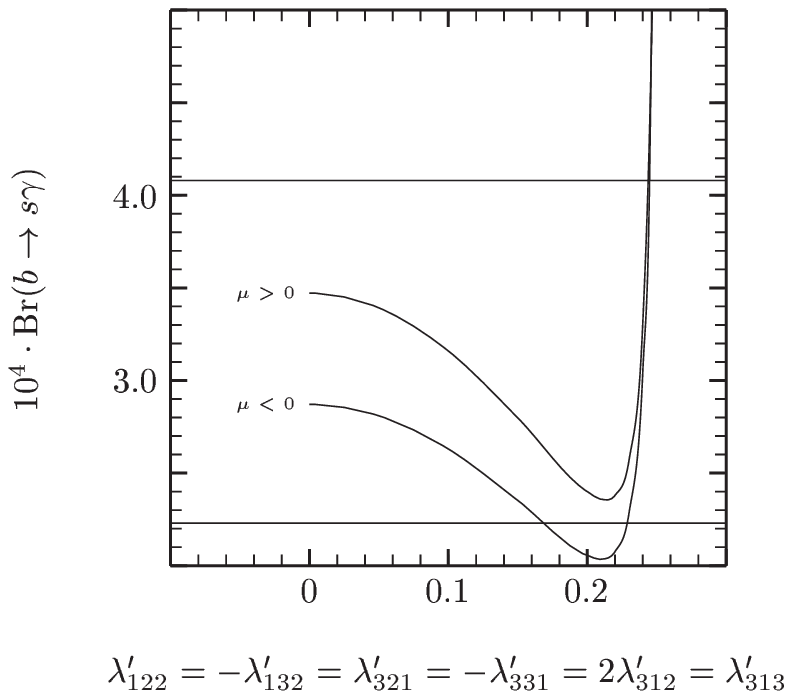}
\end{center}
\caption{\label{fig:13} Values for Br$(b\rightarrow s\gamma)$ as a function of
$-\lambda'_{132}=\lambda'_{122}=-\lambda'_{331}
=\lambda'_{321}=2\lambda'_{312}=\lambda'_{313}$ within our reference model. 
Additional 
current bounds are: $\lambda'_{331}\leq0.45 \frac{m_{\tilde{b}}}{100 \mathrm{GeV}}$,
$\lambda'_{321}\leq0.52 \frac{m_{\tilde{d}}}{100 \mathrm{GeV}}$ \protect\cite{bounds1}.}
\end{figure}

It is obvious that the Wilson coefficients depend in a very complicated way on the parameters of
our supersymmetric model. Changes of $\lambda$, $\lambda'$, or $\lambda''$ not
only affect the result in a direct way but also in an indirect fashion through
 an altered mass spectrum and different mixing
matrices. Therefore it is very hard to make general statements on the
behaviour of the branching ratio. \newline
As mentioned, it is not our aim to perform a high  precision analysis of the
parameter space \cite{pheno1,pheno2,pheno3,pheno4} but to explore the influence of the R-parity violating couplings
on $b\rightarrow s\gamma$ with a reasonable accuracy. We know that our results
for the branching ratio are only valid at the 25\% level because we do a
leading-log approximation with large scale uncertainty \cite{buras3}. However,
we expect the impact of the R-parity breaking terms not to change
much when including the next-to-leading corrections. This means that the 
shape of the curves remain more or less the same whereas the offset where our
curves start (i.e., no R-parity breaking) may change significantly when
calculating the next-to-leading-log approximation.
Solving the RGE's was performed in the following way:
\begin{itemize}
\item Solve the equations for the gauge couplings. The boundary conditions are the
physical values at $M_Z$. The gauge couplings will meet at 
$M_X\approx 2\cdot10^{16}$ GeV. Choose a common value $M(M_X)$ for the gaugino masses at that
high scale and solve the RGE's for $M_1$-$M_3$.
\item Set $\tan\beta$  and $\lambda(M_Z)$, $\lambda'(M_Z)$ and $\lambda''(M_Z)$ 
to the desired value and use them together with the quark and lepton masses as
inputs at $M_Z$ for the Yukawa couplings. Let these couplings run to $M_X$.
\item Choose $A$, $\mu_{2i}(M_X)$ and trial values for $m_0$ and $\mu(M_X)$ to
complete 
the boundary conditions at $M_X$. The whole set of RGE's is then run down to
$M_Z$.
We choose $\mu_{2i}(M_X)=0$. This is an approximation because 
$\mu_{2i}$ and $\tilde\mu_{2i}$ will not vanish at $M_Z$. However,
in our examples their values are so small compared to $\mu$ and $\tilde\mu$ that we can
neglect them avoiding a mixing between $H_1$ and $L_i$. 
\item The minimum of the
 one-loop effective Higgs potential $V_0+\Delta V$ will in general not be at
 $(v_1,v_2)$ which makes it necessary to adjust $\mu(M_X)$ and $m_0$ in a clever
 way. We chose the Newton method to converge to the desired position of the
 minimum. Following \cite{gamberini3} we evaluate the minimum of the potential
 at some average mass scale to avoid large logarithms and therefore get a more
 reliable result.
\end{itemize}
The next step is the numerical diagonalization of the mass matrices to find the
masses of the physical particles and the relevant mixing matrices.\\
Changing the value of the R-parity violating couplings makes it necessary to
continuously adjust the values of $m_0$ and $\mu(M_X)$ which alters the mass
spectrum of the particles at $M_Z$. All the following examples correspond to
mass spectra within the current bounds for the masses of the supersymmetric
particles. We encountered two critical situations, namely too small masses for
the lightest selectron and/or the lightest Higgs boson. An idea would then be to
constrain the bounds on the R-parity violating couplings through the requirement
of a phenomenologically realistic mass spectrum independent of the value for the
branching ratio Br$(b\rightarrow s\gamma)$. However, in general, a
realistic mass spectrum does not restrict the R-parity breaking parameters
substantially. Moreover, the mass spectrum highly depends on the values of
$\tan\beta$, $A$ and $M(M_X)$. Hence, such bounds would be strongly model
dependent.\\
As reference model we chose $\tan\beta=5$, $A=0$ and $M(M_X)=300$ GeV. With
vanishing R-parity violating couplings this
leads to squark masses of $600-800$ GeV, sleptons of $260-330$ GeV, a lightest
neutralino of 120 GeV and a lightest Higgs of about 100 GeV. Fig. \ref{fig:9}
shows the behaviour of Br$(b\rightarrow s\gamma)$ in the neighbourhood of our
reference model with, in addition, $\lambda'_{132}=\lambda'_{122}=0.1$.
Interestingly, the value for the branching ratio is rather stable under a change
of $\tan\beta$ and $A$.\\
As mentioned before, it is very difficult to
 isolate generic features of the different models. Let us make some
 comments:
\begin{itemize}
\item At least two of the $\lambda$'s must be non-zero to have an influence on
the result. 
There are
two exceptions: $\lambda''_{123}$ and $\lambda''_{223}$ alone will give a contribution
due to the anti-symmetry of $\lambda''$. However, their impact on the branching
ratio is so small that no reasonable bounds can be found.
\item As all the effective couplings depend on the inverse mass-squared of a
heavy particle, it is clear that the influence of the new physics is bigger in
models with a lower mass spectrum. To see the effect of smaller masses 
 compare Figs.\ \ref{fig:10} (a) and (b) where for the latter $M(M_X)=100$ GeV 
 is taken which results in, for instance, squark masses of about 250 GeV. Note,
 that our reference model leads to quite high masses. 
 \item The squarks are always heavier than the sleptons. As a consequence, the
 influence of non-vanishing $\lambda''$ on Br$(b\rightarrow s\gamma)$ is
 much smaller than for non-vanishing $\lambda'$. 
 \item The simplest non trivial situation consists of a single non-vanishing 
 pair of R-parity violating couplings. We encounter the following scenarios:
\begin{itemize}
\item The largest effect results from the pair $(\lambda'_{122},\lambda'_{132})$ 
(see Fig. \ref{fig:12}). It is not reasonable to extract a
bound for the product of these parameters because (through the RGE's) the
dependence on these couplings is much more complex. However, we draw the
conclusion that significant effects result if both couplings exceed 0.1-0.2.
Fig. \ref{fig:12} also shows that the branching ratio may strongly depend on the
relative sign of the couplings. This happens if the new contributions are able
to diminish the value of $C_7$.
\item A general feature of our model is that the influence of a pair of
non-vanishing R-parity violating couplings on the branching
ratio starts being significant if $(\lambda',\lambda')\sim 0.2-0.4$. For
$(\lambda'',\lambda'')$, in most cases, the requirement of non-diverging RGE's
for these couplings gives more stringent bounds. There is one more aspect: Some
of the contributing pairs have more than one distinct index. They appear 
only because of the mass differences of the different squark or slepton
generations. This can be seen clearly in formulas (\ref{match}): If the 
masses were independent of their index the $\Gamma$'s would combine to
an identity matrix leaving only those pairs with an identical sfermion index. The
bounds on these pairs are much less stringent. This
is exactly what happens in the case of $\lambda''_{123}$ and $\lambda''_{223}$
when being the only non-vanishing coupling. This is another reason why the 
results for this situation are not stringent.
\item A special situation is depicted in Fig.\ \ref{fig:11}. The branching
ratio seems to ``explode'' at $\lambda'_{312}=\lambda'_{313}\approx 0.3$. 
This is due to the mentioned decreasing mass squared of the lightest
selectron, which appears in some denominators in the matching conditions. Hence,
 the specific position of the peak is highly model dependent.
\end{itemize}
\item If several pairs of couplings are non-vanishing the picture can get 
more complex, as Fig.\ \ref{fig:13} shows. There, we combined the decreasing
effect with the peak due to the small selectron mass.

\item It is difficult to compare our results with existing constraints for the
R-parity breaking couplings because the results are extremely model dependent. 
However, in the case of $\lambda'$ our bounds are highly competitive. 
For comparison we put the current bounds \cite{bounds1,bounds2,bounds3}
 in the captions of the respective figures.
\end{itemize}

\section{Conclusions}

To examine the influence of R-parity breaking on $b\rightarrow s\gamma$ one has
to enlarge the operator basis substantially. At the leading-log level it consists
of 28 operators, neglecting Higgs-lepton mixing avoiding this way the
introduction of scalar ope\-rators. The corresponding $\gamma$-matrix can be found 
with the help of previously known results and diagonalized numerically. The 
matching conditions of the magnetic penguins $O_7$ and $O_8$ get new
contributions. Their counterparts of opposite chirality, $\tilde O_7$ and $\tilde
O_8$, also have to be considered. If one uses the semi-leptonic decay
$b\rightarrow u|c\;e\bar\nu$ to cancel the large bottom mass dependence new
contributions to this decay must be included.\newline
R-parity breaking definitely has influence on the branching ratio of
$b\rightarrow s\gamma$. However, its impact is highly model dependent because
the (unknown) supersymmetric masses are mostly responsible for the size of the
new contributions. In a cautious model the new couplings are able to
change significantly the result if they are of order $10^{-1}$. Moreover,  45 
new
(complex) Yukawa couplings offer an infinite number of possible scenarios. The
simplest cases involve only one or two couplings present but also in these
situations completely different evolutions of the branching ratio are possible.
To give more concrete results we definitely need more informations on the
parameters of our supersymmetric model. 

\acknowledgments
We wish to thank Francesca Borzumati for pointing out the importance of R-parity
breaking, Christoph Greub, Tobias Hurth and
especially Daniel Wyler for their support and many illuminating discussions. We also
thank the authors of \cite{chun} for pointing out to us some errors in the $\gamma$-matrix
of the previous version of this work. Fortunately, the conclusions do not undergo
significant changes.

\appendix
\section{Mixing matrices and interaction Lagrangian} \label{ami}
\subsection{mixing matrices} \cite{mssm1,mssm2}\label{am}

In this first appendix we present the mass mixing matrices for the 
relevant particles. They are needed for two reason: First, their 
eigenvalues correspond to the physical masses of the particles and 
second, the unitary matrices that diagonalize the mass matrices 
rotate 
the fields to their (physical) mass eigenstates. 
\subsubsection*{Charginos}
    The charginos $\chi^{\mathrm{ch}}_{1/2}$ are a mixture of charged 
    gauginos $\lambda^{\pm}$ and Higgsinos $h_{1}^{-}$ and 
$h_{2}^{+}$.
    Defining
    \beq
    \psi^{+}=\left(\begin{array}{c}-i\lambda^{+}\\ 
    h_{2}^{+}\end{array} \right) \hskip 2cm
    \psi^{-}=\left(\begin{array}{c}-i\lambda^{-}\\ 
    h_{1}^{-}\end{array} \right)
    \eeq
    the mass terms are then
    \beq
   {\mathcal L}_{m}^{\mathrm{ch}}=-\frac{1}{2}(\psi^{+T}X^{T}\psi^{-}+\psi^{-T}X\psi^{+})+\mbox{h.c.}, 
    \eeq
    where
    \beq
    X=\left(\begin{array}{cc}-M_{2}&g_{2}v_{2}\\ g_{2}v_{1}&-\mu 
    \end{array}\right).
    \eeq
    The two-component charginos $\chi^{\pm}_{i}\;(i=1,2)$ and the 
    four-component charginos $\chi^{\mathrm{ch}}_{1/2}$ are then defined as 
    \beq \renewcommand{\arraystretch}{1.5}
    \begin{array}{llllll}
    \chi_{i}^{+}&=&V_{ij}\psi_{j}^{+}&\chi_{i}^{-}&=&U_{ij}\psi_{j}^{-}\\
    \\ 
    \chi^{\mathrm{ch}}_{1}&=&\left(\begin{array}{c} 
    \chi_{1}^{+}\\ 
    \overline{\chi_{1}^{-}}\end{array}\right) &\chi_{2}^{\mathrm{ch}}&=& 
    \left(\begin{array}{c}\chi_{2}^{+}\\ 
    \overline{\chi_{2}^{-}}\end{array}\right),
    \end{array}
    \eeq
    where the unitary matrices $U$ and $V$ diagonalize $X$:
    \beq
    M_{D}^{\mathrm{ch}}=U^{*}XV^{-1}=VX^{\dagger}U^{T}
    \eeq
    ${\mathcal L}^{\mathrm{ch}}_{m}$ then becomes
    \beq
    {\mathcal L}^{\mathrm{ch}}_m=-M^{\mathrm{ch}}_{D\,11}\:\overline{\chi_{1}^{\mathrm{ch}}}\chi^{\mathrm{ch}}_{1} 
    -M^{\mathrm{ch}}_{D\,22}\:\overline{\chi^{\mathrm{ch}}_{2}}\chi^{\mathrm{ch}}_{2}.
    \eeq
    \\
    $U$ and $V$ can be found by observing that
    \beq
    M^{\mathrm{ch}\,2}_{D}=VX^{T}XV^{-1}=U^{*}XX^{T}U^{*-1}.
    \eeq
    They are not fixed completely by these conditions. The freedom 
can 
    be used to arrange the elements of $M^{\mathrm{ch}}_{D}$ to be positive: 
If 
    the $i^{\mathrm{th}}$ eigenvalue of $M^{\mathrm{ch}}_{D}$ is negative simply 
multiply 
    the $i^{th}$ row of $V$ with $-1$.
    
 \subsubsection*{Neutralinos}
 The neutralinos are linear combinations of the gauginos $\lambda'$ 
and 
 $\lambda_{3}$ and the neutral Higgsinos $h_{1}^{0}$ and $h_{2}^{0}$. 
 If we define
 \beq
 \psi^{0}=\left(\begin{array}{c} -i\lambda'\\-i\lambda_{3}\\h_{1}^{0} 
 \\h_{2}^{0}\end{array}\right)
 \eeq
 the neutralino mass term reads
 \beq
 {\mathcal L}^{0}_{m}=-\frac{1}{2}\psi^{0T}Y\psi^{0}+\mbox{h.c.},
 \eeq
 where
 \beq
 Y=\left(\begin{array}{cccc}
 -M_{1}&0&-\frac{g_{1}v_{1}}{\sqrt{2}}&\frac{g_{1}v_{2}}{\sqrt{2}}\\
 0&-M_{2}&\frac{g_{2}v_{1}}{\sqrt{2}}&-\frac{g_{2}v_{2}}{\sqrt{2}}\\
 -\frac{g_{1}v_{1}}{\sqrt{2}}&\frac{g_{2}v_{1}}{\sqrt{2}}&0&\mu\\
 \frac{g_{1}v_{2}}{\sqrt{2}}&-\frac{g_{2}v_{2}}{\sqrt{2}}&\mu&0
 \end{array} \right)
 \eeq
 Two- and four-component neutralinos must be defined as
 \beq \renewcommand{\arraystretch}{1.5}
 \begin{array}{l}
     \tilde{\chi}^{0}_{i}=N_{ij}\psi^{0}_{j}\hskip 1.5cm i=1,\ldots 
,4\\
     \chi^{0}_{i}=\left(\begin{array}{c}\tilde{\chi}^{0}_{i} \\ 
     \overline{\tilde \chi^{0}_{i}} \end{array}\right)
 \end{array}
\eeq 
To diagonalize the mass matrix $N$ must obey
\beq
N_{D}=N^{*}YN^{-1},
\eeq
where $N_{D}$ is a diagonal matrix.
$N$ can be found using the property
\beq
N_{D}^{2}=NY^{\dagger}YN^{-1}.
\eeq
The eigenvalues and eigenvectors are found numerically. Possible 
negative entries in $N_{D}$ are turned positive by multiplying the 
corresponding row of $N$ by a factor of $i$.
\subsubsection*{Quarks and Leptons}
The situation in the quark and lepton sector is in almost complete 
analogy to the 
standard model. The quarks and leptons get their masses from the 
Yukawa 
potential when the Higgs bosons acquire a vacuum expectation value. We 
define the mass eigenstates by
\beq
\begin{array}{ll}
    u_{Li}^{(m)}=U^{L}_{ij}u_{Lj}&u_{Ri}^{(m)}=U^{R}_{ij}u_{Rj}\\
    \\ 
    d_{Li}^{(m)}=D^{L}_{ij}d_{Lj}&d_{Ri}^{(m)}=D^{R}_{ij}d_{Rj}\\
    \\
    e_{Li}^{(m)}=E_{ij}^{L}e_{Lj}&e_{Ri}^{(m)}=E^{R}_{ij}e_{Rj}.
\end{array}
\eeq
The mixing matrices must satisfy
\beq
\begin{array}{rlllll}
    
D^{R}\lambda^{dT}D^{L\dagger}&=&\lambda^{d}_{D}&=&\mbox{diag}\left( 
    \frac{m_{di}}{v_{1}}\right)&i=1,\ldots,3\\
    \\
    
    U^{R}\lambda^{uT}U^{L\dagger}&=&\lambda^{u}_{D}&=&\mbox{diag}\left( 
    \frac{m_{ui}}{v_{2}}\right)&\\ 
    \\ 
    E^{R}\lambda^{eT}E^{L\dagger}&=&\lambda^{e}_{D}&=&\mbox{diag}\left( 
    \frac{m_{ei}}{v_{1}}\right)
\end{array}
\eeq
\\
\begin{eqnarray}
v_1&=&\sqrt{2}\frac{M_W}{g_2}\cos \beta \nonumber \\[0.5cm]
v_2&=&\sqrt{2}\frac{M_W}{g_2}\sin \beta. \label{v1v2}
\end{eqnarray}
As one can see, the eigenvalues of $\lambda^{u}$ and $\lambda^{d}$ 
are 
fixed by the quark masses and the minimum of the Higgs potential.
In the SM the only effect of the mixing which can be seen is the 
CKM-matrix $K=U^{L}D^{L\dagger}$ appearing in the flavour changing 
charged currents. Therefore it is possible and convenient to set 
\beq
D^{L}=D^{R}=U^{R}=\openone\hskip 1.5cm(\Rightarrow U^{L}=K)
\eeq
(To be more precise, one chooses $\lambda^{d}$ and $\lambda^e$ to be diagonal and 
$\lambda^{u}=K^{T}\mbox{diag}\left(m_{ui}/v_{2}\right)$.)
Although in our theory the mixing matrices appear in all kinds of 
combinations we adopt this convention here, emphasising that it is a 
\emph{choice} made just for convenience. It is possible that one day 
an 
underlying theory fixes the values of $\lambda^{u}$ and $\lambda^{d}$ 
at some (high) scale.\newline
Please note that in the text we neglect the superscript $m$ for the 
mass 
eigenstates.

\subsubsection*{Squarks and Sleptons}
If supersymmetry were not broken squarks and sleptons would be 
rotated to 
their mass basis with the help of the same matrices as their 
fermionic partners. But since this situation is not realistic we need 
to introduce a further set of unitary rotation matrices. The notation 
must be set up carefully because the mass eigenstates of squarks and 
sleptons are linear combinations of the partners of \emph{left-} and 
\emph{right-}handed partners of the corresponding fermions. We define 
the $6\times 3$-matrices $\Gamma$ in the following way:
\beq
\begin{array}{lllll}
\tilde u^{(m)}&=&\left(\Gamma^{u}_{L}|\Gamma^{u}_{R}\right) 
\left(\begin{array}{c}\tilde u_{L}\\ \tilde u_{R}\end{array}\right)
&=&\Gamma^{u}\tilde u\\
\\
\tilde d^{(m)}&=&\left(\Gamma^{d}_{L}|\Gamma^{d}_{R}\right) 
\left(\begin{array}{c}\tilde d_{L}\\ \tilde d_{R}\end{array}\right)
&=&\Gamma^{d}\tilde d\\
\\
\tilde e^{(m)}&=&\left(\Gamma^{e}_{L}|\Gamma^{e}_{R}\right) 
\left(\begin{array}{c}\tilde e_{L}\\ \tilde e_{R}\end{array}\right)
&=&\Gamma^{e}\tilde e\\
\\
\tilde \nu^{(m)}&=&\Gamma^{\nu}\tilde \nu_{L}
\end{array}
\eeq 
To diagonalize the mass terms the mixing matrices have to satisfy
\begin{eqnarray}
    \Gamma^{u}m^{2}_{\mathrm{su}}\Gamma^{u\dagger}&=&m^{2}_{\mathrm{su}D}
    \nonumber \\
    \nonumber \\
    \Gamma^{d}m^{2}_{\mathrm{sd}}\Gamma^{d\dagger}&=&m^{2}_{\mathrm{sd}D} \nonumber \\
    \nonumber \\
    \Gamma^{e}m^{2}_{\mathrm{se}}\Gamma^{e\dagger}&=&m^{2}_{\mathrm{se}D} \nonumber \\
    \nonumber \\
    \Gamma^{\nu}m^{2}_{\mathrm{s\nu}}\Gamma^{\nu\dagger}&=&m^{2}_{\mathrm{s\nu}D},
\end{eqnarray}
where the matrices on the RHS are diagonal containing the 
masses squared
of the {phy-sical} particles.\newline
The mass matrices are (with the exception of the sneutrino) of the 
form 
$\left(\begin{array}{cc}A&B\\ B^{\dagger}&C\end{array}\right)$, where
$A$, $B$ and $C$ are $3\times 3$-matrices. For the different fields 
they are (choosing $\mu$ real)
\begin{itemize}
\item up-squarks:
\begin{eqnarray}
A&=&m^{2}_{Q} +v_{2}^{2}\lambda^{u}\lambda^{u\dagger} +M^{2}_{Z}
\cos2\beta\left(\frac{1}{2}-\frac{2}{3}\sin^{2}\theta_{W}\right)\openone_{3}
\nonumber \\
B&=&\left(h^{u} +\mu\cot \beta\lambda^{u}\right)v_{2} \nonumber \\
C&=&m^{2}_{U} +v_{2}^{2}\lambda^{u\dagger}\lambda^{u} 
+\frac{2}{3}M_{Z}^{2}\cos 2\beta\sin^{2}\theta_{W}\openone_{3}
\end{eqnarray}
\item down-squarks:
\begin{eqnarray}
A&=&m^{2}_{Q} +v_{1}^{2}\lambda^{d}\lambda^{d\dagger} +M^{2}_{Z}
\cos2\beta\left(-\frac{1}{2}+\frac{1}{3}\sin^{2}\theta_{W}\right)\openone_{3}
\nonumber \\
B&=&\left(h^{d} +\mu\tan \beta\lambda^{d}\right)v_{1} \nonumber \\
C&=&m^{2}_{D} +v_{1}^{2}\lambda^{d\dagger}\lambda^{d} 
-\frac{1}{3}M_{Z}^{2}\cos 2\beta\sin^{2}\theta_{W}\openone_{3}
\end{eqnarray}
\item selectrons
\begin{eqnarray}
A&=&m^{2}_{L} +v_{1}^{2}\lambda^{e}\lambda^{e\dagger} +M^{2}_{Z}
\cos2\beta\left(-\frac{1}{2}+\sin^{2}\theta_{W}\right)\openone_{3}
\nonumber \\
B&=&\left(h^{e} +\mu\tan \beta\lambda^{e}\right)v_{1} \nonumber \\
C&=&m^{2}_{E} +v_{1}^{2}\lambda^{e\dagger}\lambda^{e} 
-M_{Z}^{2}\cos 2\beta\sin^{2}\theta_{W}\openone_{3}
\end{eqnarray}
\end{itemize}
The sneutrinos have
\beq
m^2_{\mathrm{s\nu}}=m^2_{\tilde e}+\frac{M_Z}{2}\cos 2\beta\openone_3
\eeq
\subsubsection*{Higgses}
The Higgs sector consists of two $SU(2)$ doublets $\tilde h_{1}$ and 
$\tilde h_{2}$. The real and the imaginary part of the neutral 
components mix via the matrix 
\beq 
\label{h1}
\left(
\begin{array}{cc}
    \mu^{2}+m_{H_{1}}^{2} +\frac{g_{1}^{2} +g_{2}^{2}}{4}(3v_{1}^{2} 
    -v_{2}^{2})&\tilde\mu -\frac{g_{1}^{2} +g_{2}^{2}}{2}v_{1}v_{2} \\
    \tilde\mu -\frac{g_{1}^{2} +g_{2}^{2}}{2}v_{1}v_{2} &\mu^{2} 
    +m_{H_{2}}^{2} -\frac{g_{1}^{2} +g_{2}^{2}}{4} (v_{1}^{2} 
    -3v_{2}^{2})
\end{array}
\right)
\eeq
and
\beq 
\label{h2}
\left(
\begin{array}{cc}
    \mu^{2} +m_{H_{1}}^{2} +\frac{g_{1}^{2} +g_{2}^{2}}{4} 
    (v_{1}^{2} -v_{2}^{2})&-\tilde\mu \\
    -\tilde\mu&\mu^{2} +m_{H_{2}}^{2} -\frac{g_{1}^{2} 
    +g_{2}^{2}}{4} (v_{1}^{2} -v_{2}^{2})
\end{array}
\right)
\eeq
respectively. The charged components give rise to a mass matrix of 
the 
form
\beq 
\label{h3}
\left(
\begin{array}{cc}
    \mu^{2} +m_{H_{1}}^{2} +\frac{g_{1}^{2} +g_{2}^{2}}{4} 
    (v_{1}^{2} -v_{2}^{2}) +\frac{g_{2}^{2}v_{2}^{2}}{2}&-\tilde\mu 
    +\frac{g_{2}^{2}}{2}v_{1}v_{2} \\
    -\tilde\mu +\frac{g_{2}^{2}}{2}v_{1}v_{2}&\mu^{2} +m_{H_{2}}^{2} 
    -\frac{g_{1}^{2} +g_{2}^{2}}{4} (v_{1}^{2} -v_{2}^{2}) 
    +\frac{g_{2}^{2}v_{1}^{2}}{2}
\end{array}
\right)
\eeq

All the above expressions are valid at tree-level. The vacuum 
expectation values 
$v_{1}$ and $v_{2}$ are found by  minimizing the effective Higgs 
potential. 
If one inserts the gained formulas into (\ref{h1}) - (\ref{h3}) one 
of 
the eigenvalues of (\ref{h2}) and (\ref{h3}) vanishes, indicating the 
eaten fields of the Higgs mechanism that takes place.

\subsection{Interaction Lagrangian} \label{ai}
For the evaluation of the matching conditions we need certain parts of 
the
interaction Lagrangian. In addition to equations(\ref{L1}) - 
(\ref{L3}) these
are
\subsubsection*{Squark-Quark-Chargino}
\begin{eqnarray}
{\mathcal L}_{\tilde qq\chi^{\mathrm{ch}}}&=& \tilde u_{j}\overline{d_{i}}
\left[A^d_{ij\ell}P_L +B^d_{ij\ell}P_R \right]\chi^{\mathrm{ch}\;c}_\ell 
+\tilde
u_i^\dagger
\overline{\chi^{\mathrm{ch}\;c}_\ell} \left[A^{d\dagger}_{ij\ell}P_R
+B^{d\dagger}_{ij\ell}P_L\right]d_j \nonumber \\
&&+\tilde d_j\overline{u_i}\left[A^u_{ij\ell}P_L 
+B^u_{ij\ell}P_R\right]
\chi_\ell^{\mathrm{ch}} +\tilde d^\dagger_i\overline{\chi^{\mathrm{ch}}_\ell}
\left[A^{u\dagger}_{ij\ell}P_R +B^{u\dagger}_{ij\ell}P_L\right]u_j,
\end{eqnarray}
where 
\begin{eqnarray}
A^d_{ij\ell}&=&(\lambda^d_D\Gamma^{u\dagger}_L)_{ij}U^*_{\ell 2} \nonumber \\
\nonumber \\
B^d_{ij\ell}&=&(K^\dagger\lambda^u_D\Gamma^{u\dagger}_R)_{ij}V_{\ell 
2}
-g_2\Gamma^{u\dagger}_{Lij} V_{\ell 1} \nonumber \\
\nonumber \\
A^u_{ij\ell}&=&(\lambda^u_DK\Gamma^{d\dagger}_L)_{ij}V^*_{\ell 2} \nonumber \\
\nonumber \\
B^u_{ij\ell}&=&(K\lambda^d_D\Gamma^{d\dagger}_R)_{ij}U_{\ell 2}
-g_2(K\Gamma^{d\dagger}_L)_{ij} U_{\ell 1} \nonumber \\
\nonumber \\
P_{L/R}&=&\frac{1}{2}(1\mp\gamma^5)
\end{eqnarray}
and $\chi^{\mathrm{ch}\;c}_\ell$ denotes the charge conjugated field.
\subsubsection*{Squark-Quark-Neutralino}

\begin{eqnarray}
{\mathcal L}_{\tilde qq\chi^0}&=&-\tilde 
d_j\overline{d_i}\left[C^d_{ij\ell}P_L
+D^d_{ij\ell}P_R\right]\chi^0_\ell -\tilde
d^\dagger_i\overline{\chi^0_\ell}\left[C^{d\dagger}_{ij\ell}P_R
+D^{d\dagger}_{ij\ell}P_L\right]d_j \nonumber \\
&&-\tilde u_j\overline{u_i}\left[C^u_{ij\ell}P_L
+D^u_{ij\ell}P_R\right]\chi^0_\ell
-\tilde 
u^\dagger_i\overline{\chi^0_\ell}\left[C^{u\dagger}_{ij\ell}P_R
+D^{u\dagger}_{ij\ell}P_L\right]u_j,
\end{eqnarray}
where
\begin{eqnarray}
C^d_{ij\ell}&=&(\lambda^d_D\Gamma^{d\dagger}_L)_{ij}N^*_{\ell 3}
-\sqrt{2}g_1Q_d\Gamma^{d\dagger}_{Rij}N^*_{\ell 1} \nonumber \\
\nonumber \\
D^d_{ij\ell}&=&(\lambda^d_D\Gamma^{d\dagger}_R)_{ij}N_{\ell 3}
+\frac{1}{\sqrt{2}}\Gamma^{d\dagger}_{Lij} ((2Q_d+1)g_1N_{\ell 1}-g_2 
N_{\ell
2}) \nonumber \\
\nonumber \\
C^u_{ij\ell}&=&(\lambda^u_DK\Gamma^{u\dagger}_L)_{ij}N^*_{\ell 4}
-\sqrt{2}Q_ug_1\Gamma^{u\dagger}_{Rij}N^*_{\ell 1} \nonumber \\
\nonumber \\
D^u_{ij\ell}&=&(\lambda^u_D\Gamma^{u\dagger}_R)_{ij}N_{\ell 4}
+\frac{1}{\sqrt{2}}(K\Gamma^{u\dagger}_L)_{ij}((2Q_d+1)g_1N_{\ell 1} 
+g_2N_{\ell
2}).
\end{eqnarray}
\subsubsection*{Squark-Quark-Gluino}
\begin{eqnarray}
{\mathcal L}_{\tilde qqg}&=&-\sqrt{2}g_sT^{A\alpha\beta}
\left[\tilde u^{\dagger}_{i\alpha}
\overline{g_A}\left[(\Gamma^u_LK^\dagger)_{ij}P_L
-\Gamma^u_{Rij}P_R\right]u_{j\beta} 
\right.\nonumber \\
&+&\tilde d^{\dagger}_{i\alpha} \overline{g_A}\left[\Gamma^d_{Lij}P_L 
-\Gamma^d_{Rij}P_R\right] d_{j\beta} \nonumber \\
&+&\tilde u_{i\alpha} 
\overline{u_{j\beta}}\left[(K\Gamma^{u\dagger}_L)_{ij}P_R
-\Gamma^{u\dagger}_{Rij}P_L\right]g_A \nonumber \\
&+&\left.\tilde d_{j\alpha}
\overline{d_{i\beta}}\left[\Gamma^{d\dagger}_{Lij}P_R 
-\Gamma^{d\dagger}_{Rij}P_L\right] g_A\right]
\end{eqnarray}
\subsubsection*{Gluino-Gluino-Gluon}
\beq
{\mathcal L}_{gg\tilde g}=\frac{i}{2}g_sf^{ABC}\overline{g_A}\gamma_\mu
g_BA^\mu_C
\eeq
Note: There is a symmetry factor of two in the Feynman rule for this 
vertex.

\section{Renormalization group equations} \label{ar}
We present the full set of RGE's for all the parameters of the MSSM 
including R-parity breaking terms. Our results are in complete 
agreement with \cite{decarlos2,ciuchini1}, although we don't restrict ourselves to 
couplings of the third generation. All the formulas can be derived 
from 
the expressions for the most general form of a softly broken
SUSY \cite{derendinger,gato,falck}. 
Let 
us begin with the parameters of the superpotential $W$ ($t=\ln \mu$).

\begin{eqnarray}
16\pi^{2}\frac{d}{dt}\mu&=&\mbox{Tr}\left(3\lambda^{d}\lambda^{d\dagger} 
+3\lambda^{u}\lambda^{u\dagger}+\lambda^{e}\lambda^{e\dagger}\right)\mu-\left[g_{1}^{2}+3g_{2}^{2}\right]\mu 
\nonumber \\
&&+\left[3\lambda^d_{\ell m}\lambda'^*_{i\ell m} + \lambda^e_{\ell 
m}\lambda^*_{i\ell
m}\right]\mu_{2i}\\
\nonumber \\
16\pi^{2}\frac{d}{dt}\mu_{2i}&=&\left[(\lambda^{e}\lambda^{e\dagger})_{ij}
+ \lambda_{i\ell m}\lambda_{j\ell m}^{*}+3 \lambda'_{i\ell
m}\lambda'^{*}_{j\ell m}\right]\mu_{2j} \nonumber \\
&&+ \left[3 \lambda'_{i\ell
m}\lambda^{d*}_{\ell m} + \lambda_{i\ell m}\lambda^{e*}_{\ell
m}\right]\mu \nonumber \\
&&+3\mbox{Tr}(\lambda^u\lambda^{u\dagger})\mu_{2i}-\left 
[g_1^2+3g_2^2\right]
\mu_{2i} \\
\nonumber \\ 
16\pi^{2}\frac{d}{dt}\lambda^{u}_{ij}&=&\left[(\lambda^{u}\lambda^{u\dagger})_{ik} 
+(\lambda^{d}\lambda^{d\dagger})_{ik} +\lambda'^{*}_{\ell k 
m}\lambda'_{\ell i m}\right]\lambda^{u}_{kj} \nonumber \\ 
&&+\left[2(\lambda^{u\dagger}\lambda^{u})_{kj} +\lambda''^{*}_{k\ell m 
}\lambda''_{j\ell m}\right]\lambda^{u}_{ik} \nonumber \\
&&+3\mbox{Tr}(\lambda^{u}\lambda^{u\dagger})\lambda^{u}_{ij} 
-\left[\frac{13}{9}g_{1}^{2}+3g_{2}^{2}+\frac{16}{3}g_{3}^{2}\right] 
\lambda^{u}_{ij}\\
\nonumber \\
16\pi^{2}\frac{d}{dt}\lambda^{d}_{ij}&=&\left[(\lambda^{u}\lambda^{u\dagger})_{ik} 
+(\lambda^{d}\lambda^{d\dagger})_{ik} +\lambda'^{*}_{\ell k 
m}\lambda'_{\ell i m}\right]\lambda^{d}_{kj} \nonumber \\ 
&&+\left[2(\lambda^{d\dagger}\lambda^{d})_{kj} +2\lambda'^{*}_{\ell m 
k} \lambda'_{\ell m j} +2 \lambda''^{*}_{\ell k m}\lambda''_{\ell j 
m}\right]\lambda^{d}_{ik} \nonumber \\
&&+\left[3\mbox{Tr}(\lambda^{d}\lambda^{d\dagger})+\mbox{Tr}(\lambda^{e} 
\lambda^{e\dagger})\right]\lambda^{d}_{ij} \nonumber \\
&&+\left[3\lambda^{d}_{\ell m}\lambda'^{*}_{k\ell 
m}+\lambda^{e}_{\ell 
m}\lambda^{*}_{k\ell m}\right]\lambda'_{kij}\nonumber \\
&&-\left[\frac{7}{9}g_{1}^{2}+3 
g_{2}^{2}+\frac{16}{3}g_{3}^{2}\right]\lambda^{d}_{ij} \\
\nonumber \\
16\pi^{2}\frac{d}{dt}\lambda^{e}_{ij}&=&\left[(\lambda^{e}\lambda^{e\dagger})_{ik} 
+\lambda^{*}_{k\ell m}\lambda_{i\ell m} +3\lambda'^{*}_{k\ell 
m}\lambda'_{i\ell m}\right]\lambda^{e}_{kj} \nonumber \\
&&+\left[2(\lambda^{e\dagger}\lambda^{e})_{kj} +\lambda^{*}_{\ell m 
k}\lambda_{\ell m j}\right]\lambda^{e}_{ik} \nonumber \\
&&+\left[3\mbox{Tr}(\lambda^{d}\lambda^{d\dagger}) 
+\mbox{Tr}(\lambda^{e}\lambda^{e\dagger})\right]\lambda^{e}_{ij} 
\nonumber \\
&&+\frac{1}{2}\left[3\lambda^{d}_{\ell m}\lambda'^{*}_{k\ell m} 
+\lambda^{e}_{\ell m}\lambda^{*}_{k\ell m}\right]\lambda_{kij} \nonumber \\
&&-\left[3g_{1}^{2}+3g_{2}^{2}\right]\lambda^{e}_{ij}\\
\nonumber \\
16\pi^{2}\frac{d}{dt}\lambda_{ijk}&=&\left[(\lambda^{e}\lambda^{e\dagger})_{in} 
+\lambda^{*}_{n\ell m}\lambda_{i\ell m} +3\lambda'^{*}_{n\ell 
m}\lambda'_{i\ell m}\right]\lambda_{njk} \nonumber \\
&&+\left[(\lambda^{e}\lambda^{e\dagger})_{jn} 
+\lambda^{*}_{n\ell m}\lambda_{j\ell m} +3\lambda'^{*}_{n\ell 
m}\lambda'_{j\ell m}\right]\lambda_{ink} \nonumber \\
&&+\left[2(\lambda^{e\dagger}\lambda^{e})_{nk} + \lambda^{*}_{\ell m 
n}\lambda_{\ell m k}\right]\lambda_{ijn} \nonumber \\ 
&&+2\left[3\lambda^{d*}_{\ell m}\lambda'_{i\ell m} 
+\lambda^{e*}_{\ell 
m}\lambda_{i\ell m}\right]\lambda^{e}_{jk} \nonumber \\
&&-2\left[3\lambda^{d*}_{\ell m}\lambda'_{j\ell m} 
+\lambda^{e*}_{\ell 
m}\lambda_{j\ell m}\right]\lambda^{e}_{ik} \nonumber \\
&&-\left[3g_{1}^{2}+3g_{2}^{2}\right]\lambda_{ijk}\\
\nonumber \\
16\pi^{2}\frac{d}{dt}\lambda'_{ijk}&=&\left[(\lambda^{e}\lambda^{e\dagger})_{in} 
+\lambda^{*}_{n\ell m}\lambda_{i\ell m} +3\lambda'^{*}_{n\ell 
m}\lambda'_{i\ell m}\right]\lambda'_{njk}\nonumber \\
&&+\left[(\lambda^{u}\lambda^{u\dagger})_{jn} 
+(\lambda^{d}\lambda^{d\dagger})_{jn} +\lambda'^{*}_{\ell n 
m}\lambda'_{\ell j m}\right]\lambda'_{ink} \nonumber \\
&&+\left[2(\lambda^{d\dagger}\lambda^{d})_{nk} +2\lambda'^{*}_{\ell m 
n}\lambda'_{\ell m k} +2\lambda''^{*}_{\ell n m}\lambda''_{\ell 
km}\right] \lambda'_{ijn} \nonumber \\
&&+\left[3\lambda^{d*}_{\ell m}\lambda'_{i\ell m} 
+\lambda^{e*}_{\ell m}\lambda_{i\ell m} \right]\lambda^{d}_{jk} 
\nonumber \\
&&-\left[\frac{7}{9}g_{1}^{2}+3g_{2}^{2}+\frac{16}{3}g_{3}^{2}\right] 
\lambda'_{ijk} \\
\nonumber \\
16\pi^{2}\frac{d}{dt}\lambda''_{ijk}&=&\left[2(\lambda^{u\dagger}\lambda^{u})_{ni} 
+\lambda''^{*}_{n\ell m}\lambda''_{i\ell m}\right]\lambda''_{njk} 
\nonumber \\
&&+\left[2(\lambda^{d\dagger}\lambda^{d})_{nj} +2\lambda'^{*}_{\ell 
n 
m}\lambda'_{\ell m j} +2\lambda''^{*}_{\ell n m}\lambda''_{\ell j m} 
\right]\lambda''_{i n k} \nonumber \\
&&+\left[2(\lambda^{d\dagger}\lambda^{d})_{nk} +2\lambda'^{*}_{\ell 
n 
m}\lambda'_{\ell m k} +2\lambda''^{*}_{\ell n m}\lambda''_{\ell k m} 
\right]\lambda''_{i j n} \nonumber \\
&&-\left[\frac{4}{3}g_{1}^{2}+8g_{3}^{2}\right]\lambda''_{ijk}
\end{eqnarray}
\\
The parameters of $V_{\mathrm{soft}}$ obey the following RGE's:

\begin{eqnarray}
16\pi^2\frac{d}{dt}\tilde\mu&=&\mbox{Tr}\left(3\lambda^d\lambda^{d\dagger}
+3\lambda^u\lambda^{u\dagger} 
+\lambda^e\lambda^{e\dagger}\right)\tilde\mu
\nonumber \\
&&+\left[3\lambda^d_{\ell m}\lambda'^*_{i\ell m} +\lambda^e_{\ell m}
\lambda^*_{i\ell m}\right]\tilde\mu_{2i} \nonumber \\
&&+2\mbox{Tr}\left(3h^d\lambda^{d\dagger} +3h^u\lambda^{u\dagger}
+h^e\lambda^{e\dagger}\right)\mu \nonumber \\
&&+2\left[3h^d_{\ell m}\lambda'^*_{i\ell m} +h^e_{\ell 
m}\lambda^*_{i\ell
m}\right]\mu_{2i} \nonumber \\
&&-\left[g_1^2+3g_2^2\right]\tilde\mu 
+\left[2M_1g_1^2+6M_2g_2^2\right]\mu \\
\nonumber \\
16\pi^2\frac{d}{dt}\tilde\mu_{2i}&=&\left[(\lambda^e\lambda^{e\dagger})_{ij}
+\lambda^*_{j\ell m}\lambda_{i\ell m} +3\lambda'^*_{j\ell 
m}\lambda'_{i\ell m}
\right]\tilde\mu_{2j} 
+3\mbox{Tr}(\lambda^u\lambda^{u\dagger})\tilde\mu_{2i}
\nonumber \\
&&+\left[3\lambda^{d*}_{\ell m}\lambda'_{i\ell m} 
+\lambda^{e*}_{\ell
m}\lambda_{i\ell m}\right]\tilde\mu \nonumber \\
&&+2\left[(h^e\lambda^{e\dagger})_{ij} +C_{i\ell m}\lambda^*_{j\ell m}
+3C'_{i\ell m}\lambda'^*_{j\ell m}\right]\mu_{2j} \nonumber \\
&&+6\mbox{Tr}(h^u\lambda^{u\dagger})\mu_{2i}
+2\left[3\lambda^{d*}_{\ell m}C'_{i\ell m} +\lambda^{e*}_{\ell 
m}C_{i\ell m}\right]\mu \nonumber \\
&&-\left[g_1^2+3g_2^2\right]\tilde\mu_{2i}
+\left[2M_1g_1^2 +6M_2g_2^2\right]\mu_{2i} \\
\nonumber \\
16\pi^2\frac{d}{dt}h^u_{ij}&=&5(\lambda^u\lambda^{u\dagger}h^u)_{ij}
+(\lambda^d\lambda^{d\dagger}h^u)_{ij} 
+4(h^u\lambda^{u\dagger}\lambda^u)_{ij}
\nonumber \\
&&+2(h^d\lambda^{d\dagger}\lambda^u)_{ij}
+3\mbox{Tr}(\lambda^u\lambda^{u\dagger})h^u_{ij}
+6\mbox{Tr}(h^u\lambda^{u\dagger})\lambda^u_{ij} \nonumber \\
&&+\lambda'^*_{m\ell n}\lambda'_{min}h^u_{\ell j} +\lambda''^*_{\ell
nm}\lambda''_{jnm}h^u_{i\ell} 
+2\lambda'^*_{mn\ell}C'_{mi\ell}\lambda^u_{nj} \nonumber \\
&&+2\lambda''^*_{\ell mn}C''_{jmn}\lambda^u_{i\ell} 
+\frac{13}{9}g_1^2(2M_1\lambda^u_{ij}-h^u_{ij}) \nonumber \\
&&+3g_2^2(2M_2\lambda^u_{ij}-h^u_{ij}) 
+\frac{16}{3}g_3^2(2M_3\lambda^u_{ij}-h^u_{ij}) \\
\nonumber \\
16\pi^2\frac{d}{dt}h^d_{ij}&=&5(\lambda^d\lambda^{d\dagger}h^{d})_{ij}
+(\lambda^u\lambda^{u\dagger}h^d)_{ij} 
+4(h^d\lambda^{d\dagger}\lambda^d)_{ij}
\nonumber \\
&&+2(h^u\lambda^{u\dagger}\lambda^d)_{ij} 
+\mbox{Tr}(3\lambda^d\lambda^{d\dagger}
+\lambda^e\lambda^{e\dagger})h^d_{ij} \nonumber \\
&&+\mbox{Tr}(6h^d\lambda^{d\dagger}
+2h^e\lambda^{e\dagger})\lambda^{d}_{ij} \nonumber \\
&&+\left[\lambda^e_{\ell m}C'_{kij} +2h^e_{\ell
m}\lambda'_{kij}\right]\lambda^*_{k\ell m} \nonumber \\
&&+3\left[\lambda^d_{\ell m}C'_{kij}
+2h^d_{\ell m}\lambda'_{kij}\right]\lambda'^*_{k\ell m} \nonumber \\
&&+\left[h^d_{kj}\lambda'_{\ell im} +2\lambda^d_{kj}C'_{\ell
im}\right]\lambda'^*_{\ell km} \nonumber \\
&&+2\left[h^d_{ik}\lambda'_{\ell mj} +2\lambda^d_{ik}C'_{\ell
mj}\right]\lambda'^*_{\ell mk} \nonumber \\
&&+2\left[h^d_{ik}\lambda''_{\ell jm}
+2\lambda^d_{ik}C''_{\ell jm}\right]\lambda''^*_{\ell km} \nonumber \\
&&+\frac{7}{9}g_1^2(2M_1\lambda^d_{ij} -h^d_{ij}) 
+3g_2^2(2M_2\lambda^d_{ij}
-h^d_{ij}) \nonumber \\
&&+\frac{16}{3}g_3^2(2M_3\lambda^d_{ij} -h^d_{ij}) \\
\nonumber \\
16\pi^2\frac{d}{dt}h^e_{ij}&=&5(\lambda^e\lambda^{e\dagger}h^e)_{ij}
+4(h^e\lambda^{e\dagger}\lambda^e)_{ij} \nonumber \\
&&+\mbox{Tr}(3\lambda^d\lambda^{d\dagger} 
+\lambda^e\lambda^{e\dagger})h^e_{ij}
+\mbox{Tr}(6h^d\lambda^{d\dagger} 
+2h^e\lambda^{e\dagger})\lambda^e_{ij}
\nonumber \\
&&+\left[\lambda^e_{\ell m}C_{kij} +2h^e_{\ell
m}\lambda_{kij} +h^e_{kj}\lambda_{i\ell m} +2\lambda^e_{kj}C_{i\ell
m}\right]\lambda^*_{k\ell m} \nonumber \\
&&+3\left[\lambda^d_{\ell m}C_{kij} +2h^d_{\ell m}\lambda_{kij}
+h^e_{kj}\lambda'_{i\ell m} +2\lambda^e_{kj}C'_{i\ell 
m}\right]\lambda'^*_{k\ell
m} \nonumber \\
&&+\left[h^e_{ik}\lambda_{\ell mj} +2\lambda^e_{ik}C_{\ell
mj}\right]\lambda^*_{\ell mk} \nonumber \\
&&+3g_1^2(2M_1\lambda^e_{ij}-h^e_{ij}) +3g_2^2(2M_2\lambda^e_{ij} 
-h^e_{ij}) \\
\nonumber \\
16\pi^2\frac{d}{dt}C_{ijk}&=&\left[(\lambda^e\lambda^{e\dagger})_{in}
+\lambda^*_{n\ell m}\lambda_{i\ell m} +3\lambda'^*_{n\ell 
m}\lambda'_{i\ell
m}\right]C_{njk} \nonumber \\
&&+\left[(\lambda^e\lambda^{e\dagger})_{jn} +\lambda^*_{n\ell 
m}\lambda_{j\ell m}
+3\lambda'^*_{n\ell m}\lambda'_{j\ell m}\right]C_{ink} \nonumber \\
&&+\left[3\lambda^{d*}_{\ell m}\lambda'_{i\ell m} +\lambda^{e*}_{\ell 
m}\lambda_{i\ell
m}\right]h^e_{jk} -\left[3\lambda^{d*}_{\ell m}\lambda'_{j\ell m}
+\lambda^{e*}_{\ell m}\lambda_{j\ell m}\right]h^e_{ik} \nonumber \\
&&+\left[2(\lambda^{e\dagger}\lambda^e)_{nk} +\lambda^*_{\ell m 
n}\lambda_{\ell
mk}\right]C_{ijn} \nonumber \\
&&+\left[2(h^e\lambda^{e\dagger})_{in} +2\lambda^*_{n\ell m}C_{i\ell 
m}
+6\lambda'^*_{n\ell m}C'_{i\ell m}\right]\lambda_{njk} \nonumber \\
&&+\left[2(h^e\lambda^{e\dagger})_{jn} +2\lambda^*_{n\ell m}C_{j\ell 
m}
+6\lambda'^*_{n\ell m}C'_{j\ell m}\right]\lambda_{ink} \nonumber \\
&&+\left[6\lambda^{d*}_{\ell m}C'_{i\ell m} +2\lambda^{e*}_{\ell 
m}C_{i\ell
m}\right]\lambda^e_{jk} \nonumber \\
&&-\left[6\lambda^{d*}_{\ell m}C'_{j\ell m}
+2\lambda^{e*}_{\ell m}C_{j\ell m}\right]\lambda^e_{ik} \nonumber \\
&&+\left[4(\lambda^{e\dagger}h^e)_{nk} +2\lambda^*_{\ell mn}C_{\ell
mk}\right]\lambda_{ijn} \nonumber \\
&&+3g_1^2(2M_1\lambda_{ijk}-C_{ijk}) 
+3g_2^2(2M_2\lambda_{ijk}-C_{ijk}) \\
\nonumber \\
16\pi^2\frac{d}{dt}C'_{ijk}&=&\left[(\lambda^e\lambda^{e\dagger})_{in}
+\lambda^*_{n\ell m}\lambda_{i\ell m} +3\lambda'^*_{n\ell 
m}\lambda'_{i\ell
m}\right]C'_{njk} \nonumber \\
&&+\left[(\lambda^u\lambda^{u\dagger})_{jn} 
+(\lambda^d\lambda^{d\dagger})_{jn}
+\lambda'^*_{\ell nm}\lambda'_{\ell jm}\right]C'_{ink} \nonumber \\
&&+\left[2(\lambda^{d\dagger}\lambda^d)_{nk} +2\lambda'^*_{\ell
mn}\lambda'_{\ell mk} +2\lambda''^*_{\ell nm}\lambda''_{\ell 
km}\right]C'_{ijn}
\nonumber \\
&&+\left[3\lambda^{d*}_{\ell m}\lambda'^*_{i\ell m} 
+\lambda^{e*}_{\ell
m}\lambda_{i\ell m}\right]h^d_{jk} \nonumber \\
&&+\left[2(h^e\lambda^{e\dagger})_{in} +2\lambda^*_{n\ell m}C_{i\ell 
m}
+6\lambda'^*_{n\ell m}C'_{i\ell m}\right]\lambda'_{njk} \nonumber \\
&&+\left[2(h^u\lambda^{u\dagger})_{jn} +2(h^d\lambda^{d\dagger})_{jn}
+2\lambda'^*_{\ell nm}C'_{\ell jm}\right]\lambda'_{ink} \nonumber \\
&&+\left[4(\lambda^{d\dagger}h^d)_{nk} +4\lambda'^*_{\ell mn}C'_{\ell 
mk}
+4\lambda''^*_{\ell nm}C''_{\ell km}\right]\lambda'_{ijn} \nonumber \\
&&+\left[6\lambda^{d*}_{\ell m}C'_{i\ell m} +2\lambda^{e*}_{\ell 
m}C_{i\ell
m}\right]\lambda^d_{jk} \nonumber \\
&&+\frac{7}{9}g_1^2(2M_1\lambda'_{ijk}-C'_{ijk})
+3g_2^2(2M_2\lambda'_{ijk}-C'_{ijk}) \nonumber \\
&&+\frac{16}{3}g_3^2(2 
M_3\lambda'_{ijk}-C'_{ijk}) \\
\nonumber \\
16\pi^2\frac{d}{dt}C''_{ijk}&=&\left[2(\lambda^{u\dagger}\lambda^u)_{ni}
+\lambda''^*_{n\ell m}\lambda''_{i\ell m}\right]C''_{njk} \nonumber \\
&&+\left[2(\lambda^{d\dagger}\lambda^d)_{nj} +2\lambda'^*_{\ell
mn}\lambda'_{\ell mj} +2\lambda''^*_{\ell nm}\lambda''_{\ell 
jm}\right]C''_{ink}
\nonumber \\
&&+\left[2(\lambda^{d\dagger}\lambda^d)_{nk} +2\lambda'^*_{\ell
mn}\lambda'_{\ell mk} +2\lambda''^*_{\ell nm}\lambda''_{\ell 
km}\right]C''_{ijn}
\nonumber \\
&&+\left[4(\lambda^{u\dagger}h^u)_{ni} +2\lambda''^*_{n\ell 
m}C''_{i\ell
m}\right]\lambda''_{njk} \nonumber \\
&&+\left[4(\lambda^{d\dagger}h^d)_{nj} +4\lambda'^*_{\ell mn}C'_{\ell 
mj}
+4\lambda''^*_{\ell nm}C''_{\ell jm}\right]\lambda''_{ink} \nonumber 
\\
&&+\left[4(\lambda^{d\dagger} h^d)_{nk} +4\lambda'^*_{\ell 
mn}C'_{\ell mk}
+4\lambda''^*_{\ell nm}C''_{\ell km}\right]\lambda''_{ijn} \nonumber 
\\
&&+\frac{4}{3}g_1^2(2M_1\lambda''_{ijk}-C''_{ijk})
+8g_3^2(2M_3\lambda''_{ijk}-C''_{ijk}) \\
\nonumber \\
16\pi^2\frac{d}{dt}m^2_{U_iU_j}&=&\left[2(\lambda^{u\dagger}\lambda^u)_{ki}
+\lambda''^*_{k\ell m}\lambda''_{i\ell m}\right]m^2_{U_kU_j} \nonumber \\
&&+\left[2(\lambda^{u\dagger}\lambda^u)_{jk} +\lambda''^*_{j\ell 
m}\lambda''_{k\ell
m}\right]m^2_{U_iU_k} \nonumber \\
&&+4\lambda^{u\dagger}_{j\ell}\lambda^u_{mi}m^2_{Q_\ell Q_m}
+4\lambda^{u\dagger}_{j\ell}\lambda^u_{\ell i}m^2_{H_2H_2} 
+4\lambda''^*_{j\ell
m}\lambda''_{inm}m^2_{D_\ell D_n} \nonumber \\
&&+4(h^{u\dagger}h^u)_{ji} +2C''^*_{j\ell m}C''_{i\ell m}
-\frac{32}{9}g_1^2M_1^2\delta_{ij}-\frac{32}{3}g_3^2M_3^2\delta_{ij} 
\nonumber \\ 
\\
16\pi^2\frac{d}{dt}m^2_{D_iD_j}&=&\left[2(\lambda^{d\dagger}\lambda^d)_{jk}
+2\lambda'^*_{\ell mj}\lambda'_{\ell mk} +2\lambda''^*_{\ell 
jm}\lambda''_{\ell
km}\right]m^2_{D_iD_k} \nonumber \\
&&+\left[2(\lambda^{d\dagger}\lambda^d)_{ki}
+2\lambda'^*_{\ell mk}\lambda'_{\ell mi} +2\lambda''^*_{\ell 
km}\lambda''_{\ell
im}\right]m^2_{D_kD_j} \nonumber \\
&&+4\lambda^{d\dagger}_{j\ell}\lambda^d_{\ell i}m^2_{H_1H_1}
+4\lambda^{d\dagger}_{j\ell}\lambda^d_{mi}m^2_{Q_\ell Q_m} 
+4\lambda'^*_{\ell
nj}\lambda'_{mni}m^2_{L_\ell L_m} \nonumber \\
&&+4\lambda'^*_{n\ell j}\lambda'_{nmi}m^2_{Q_\ell Q_m} 
+4\lambda''^*_{n\ell
j}\lambda''_{nmi}m^2_{D_\ell D_m} +4\lambda''^*_{\ell
nj}\lambda''_{mni}m^2_{U_\ell U_m} \nonumber \\
&&+4\lambda'^*_{knj}\lambda^d_{ni}m^2_{H_1L_k} 
+4\lambda^{d*}_{nj}\lambda'_{kni} m^{2}_{L_{k}H_{1}}
\nonumber \\
&&+4(h^{d\dagger}h^d)_{ji} +4C'^*_{\ell mj}C'_{\ell mi} +4C''^*_{\ell
jm}C''_{\ell im} \nonumber \\
&&-\frac{8}{9}g_1^2M_1^2\delta_{ij}-\frac{32}{3}g_3^2M_3^2\delta_{ij} 
\\
\nonumber \\
16\pi^2\frac{d}{dt}m^2_{Q_iQ_j}&=&\left[(\lambda^u\lambda^{u\dagger})_{kj}
+(\lambda^d\lambda^{d\dagger})_{kj} +\lambda'^*_{\ell 
jm}\lambda'_{\ell
km}\right]m^2_{Q_iQ_k} \nonumber \\
&&+\left[(\lambda^u\lambda^{u\dagger})_{ik} 
+(\lambda^d\lambda^{d\dagger})_{ik}
+\lambda'^*_{\ell km}\lambda'_{\ell im}\right]m^2_{Q_kQ_j} \nonumber 
\\
&&+2\lambda^{u\dagger}_{\ell j}\lambda^u_{i\ell}m^2_{H_2H_2}
+2\lambda^{u\dagger}_{mj}\lambda^u_{i\ell}m^2_{U_mU_\ell}
+2\lambda^{d\dagger}_{\ell j}\lambda^d_{i\ell}m^2_{H_1H_1} \nonumber 
\\
&&+2\lambda^{d\dagger}_{mj}\lambda^d_{i\ell}m^2_{D_mD_\ell}
+2\lambda'^*_{mjn}\lambda'_{\ell in}m^2_{L_mL_\ell}
+2\lambda'^*_{njm}\lambda'_{ni\ell}m^2_{D_mD_\ell} \nonumber \\
&&+2\lambda'^*_{kjn}\lambda^d_{in}m^2_{L_kH_1}
+2\lambda^{d*}_{jn}\lambda'_{kin}m^2_{H_1L_k} \nonumber \\
&&+2(h^uh^{u\dagger})_{ij} +2(h^dh^{d\dagger})_{ij} +2C'^*_{\ell 
jm}C'_{\ell im}
\nonumber \\
&&-\frac{2}{9}g_1^2M_1^2\delta_{ij} -6g_2^2M_2^2\delta_{ij}
-\frac{32}{3}g_3^2M_3^2\delta_{ij} \\
\nonumber \\
16\pi^2\frac{d}{dt}m^2_{E_iE_j}&=&\left[2(\lambda^{e\dagger}\lambda^e)_{jk}
+\lambda^*_{\ell mj}\lambda_{\ell mk}\right]m^2_{E_iE_k} \nonumber \\
&&+\left[2(\lambda^{e\dagger}\lambda^e)_{ki} +\lambda^*_{\ell 
mk}\lambda_{\ell
mi}\right]m^2_{E_kE_j} \nonumber \\
&&+4\lambda^{e\dagger}_{jk}\lambda^e_{ki}m^2_{H_1H_1}
+4\lambda^{e\dagger}_{jm}\lambda^e_{\ell i}m^2_{L_mL_\ell}
+4\lambda^*_{nmj}\lambda_{n\ell i}m^2_{L_mL_\ell} \nonumber \\
&&+4\lambda^*_{nkj}\lambda^e_{ni}m^2_{L_kH_1}
+4\lambda^{e*}_{nj}\lambda_{nki}m^2_{H_1L_k} \nonumber \\
&&+4(h^{e\dagger}h^e)_{ij} +2C^*_{\ell mj}C_{\ell mi} 
-8g_1^2M_1^2\delta_{ij} \\
\nonumber \\
16\pi^2\frac{d}{dt}m^2_{L_iL_j}&=&\left[(\lambda^e\lambda^{e\dagger})_{ik}
+\lambda^*_{k\ell m}\lambda_{i\ell m} +3\lambda'^*_{k\ell 
m}\lambda'_{i\ell
m}\right]m^2_{L_kL_j} \nonumber \\
&&+\left[(\lambda^e\lambda^{e\dagger})_{kj} +\lambda^*_{j\ell 
m}\lambda_{k\ell
m} +3\lambda'^*_{j\ell m}\lambda'_{k\ell m}\right]m^2_{L_iL_k} 
\nonumber \\
&&+\left[3\lambda^{d*}_{\ell m}\lambda'_{i\ell m} +\lambda^{e*}_{\ell
m}\lambda_{i\ell m}\right]m^2_{H_1L_j} \nonumber \\
&&+\left[3\lambda'^*_{j\ell m}\lambda^d_{\ell m} +\lambda^*_{j\ell
m}\lambda^e_{\ell m}\right]m^2_{L_iH_1} \nonumber \\
&&+2\lambda^{e\dagger}_{kj}\lambda^e_{ik}m^2_{H_1H_1}
+2\lambda^{e\dagger}_{mj}\lambda^e_{i\ell}m^2_{E_mE_\ell}
+2\lambda^*_{mjn}\lambda_{\ell in}m^2_{L_mL_\ell} \nonumber \\
&&+2\lambda^*_{njm}\lambda_{ni\ell}m^2_{E_mE_\ell}
+6\lambda'^*_{jmn}\lambda'_{i\ell n}m^2_{Q_mQ_\ell}
+6\lambda'^*_{jnm}\lambda'_{in\ell}m^2_{D_mD_\ell} \nonumber \\
&&+2\lambda^*_{kjn}\lambda^e_{in}m^2_{L_kH_1}
+2\lambda^{e*}_{jn}\lambda_{kin}m^2_{H_1L_k} \nonumber \\
&&+2(h^eh^{e\dagger})_{ij} +2C^*_{j\ell m}C_{i\ell m} +6C'^*_{j\ell 
m}C'_{i\ell
m} \nonumber \\
&&-2g_1^2M_1^2\delta_{ij} -6g_2^2M_2^2\delta_{ij} \\
\nonumber \\
16\pi^2\frac{d}{dt}m^2_{H_1H_1}&=&\mbox{Tr}(6\lambda^d\lambda^{d\dagger}
+2\lambda^e\lambda^{e\dagger})m^2_{H_1H_1} \nonumber \\
&&+\left[3\lambda'^*_{i\ell
m}\lambda^d_{\ell m} +\lambda^*_{i\ell m}\lambda^e_{\ell 
m}\right]m^2_{L_iH_1}
\nonumber \\
&&+\left[3\lambda^{d*}_{\ell m}\lambda'_{i\ell m} +\lambda^{e*}_{\ell
m}\lambda_{i\ell m}\right]m^2_{H_1L_i} \nonumber \\
&&+6\lambda^{d\dagger}_{nm}\lambda^d_{\ell n}m^2_{Q_mQ_\ell}
+6\lambda^{d\dagger}_{mn}\lambda^d_{n\ell}m^2_{D_mD_\ell} \nonumber \\
&&+2\lambda^{e\dagger}_{nm}\lambda^e_{\ell n}m^2_{L_mL_\ell}
+2\lambda^{e\dagger}_{mn}\lambda^e_{n\ell}m^2_{E_mE_\ell} \nonumber \\
&&+\mbox{Tr}(6h^dh^{d\dagger} +2h^eh^{e\dagger}) \nonumber \\
&&-2g_1^2M_1^2 -6g_2^2M_2^2 \\
\nonumber \\
16\pi^2\frac{d}{dt}m^2_{L_iH_1}&=&\left[(\lambda^e\lambda^{e\dagger})_{ik}
+\lambda^*_{k\ell m}\lambda_{i\ell m} +3\lambda'^*_{k\ell 
m}\lambda'_{i\ell
m}\right]m^2_{L_kH_1} \nonumber \\
&&+\mbox{Tr}(3\lambda^d\lambda^{d\dagger}
+\lambda^e\lambda^{e\dagger})m^2_{L_iH_1} \nonumber \\
&&+\left[3\lambda^{d*}_{\ell
m}\lambda'_{i\ell m} +\lambda^{e*}_{\ell m}\lambda_{i\ell 
m}\right]m^2_{H_1H_1}
\nonumber \\
&&+\left[3\lambda^{d*}_{\ell m}\lambda'_{k\ell m} +\lambda^{e*}_{\ell
m}\lambda_{k\ell m}\right]m^2_{L_iL_k} \nonumber \\
&&+6\lambda^{d*}_{nm}\lambda'_{in\ell}m^2_{D_mD_\ell}
+6\lambda^{d*}_{mn}\lambda'_{i\ell n}m^2_{Q_mQ_\ell} \nonumber \\
&&+2\lambda^{e*}_{nm}\lambda_{ni\ell}m^2_{E_mE_\ell}
+2\lambda^{e*}_{mn}\lambda_{\ell in}m^2_{L_mL_\ell} \nonumber \\
&&+6h^{d*}_{\ell m}C'_{i\ell m} +2h^{e*}_{\ell m}C_{i\ell m} \\
\nonumber \\
16\pi^2\frac{d}{dt}m^2_{H_2H_2}&=&6\mbox{Tr}(\lambda^u\lambda^{u\dagger})
m^2_{H_2H_2} \nonumber \\
&&+6\lambda^{u\dagger}_{nm}\lambda^u_{\ell n}m^2_{Q_mQ_\ell}
+6\lambda^{u\dagger}_{mn}\lambda^u_{n\ell}m^2_{U_mU_\ell} \nonumber \\
&&+6\mbox{Tr}(h^uh^{u\dagger}) -2g_1^2M_1^2 
-6g_2^2M_2^2
\end{eqnarray}
\\
Gauge couplings and gaugino masses run in the following way

\beq
\begin{array}{lcllcl}
16\pi^2\frac{d}{dt}g_1&=&11g_1^3&\hskip 1cm 
16\pi^2\frac{d}{dt}M_1&=&22g_1^2M_1\\
\\
16\pi^2\frac{d}{dt}g_2&=&g_2^3&\hskip 1cm 
16\pi^2\frac{d}{dt}M_2&=&2g_2^2M_2\\
\\
16\pi^2\frac{d}{dt}g_3&=&-3g_3^3&\hskip 1cm 
16\pi^2\frac{d}{dt}M_3&=&-6g_3^2M_3\\
\end{array}
\eeq

\section{The $\gamma$-matrix} \label{agamma}

The QCD-mixing of our new operator basis leads to a $28\times 
28$-matrix. How
its elements are deduced is explained in section \ref{gamma}. 
Fortunately, many
entries vanish giving us a chance to derive the eigenvalues and 
-vectors with
the help of \emph{Mathematica}. We split $\gamma^{{\mathrm{eff}}}$ 
in the
three block mentioned in section \ref{gamma}. The four-fermion 
operators give
the block (we have included the ordering of the operators in the 
first row)
\linespread{1}
\beq
\footnotesize{
\renewcommand{\arraystretch}{1.5}
\setlength{\arraycolsep}{0.02cm}
\left(
\begin{array}{cccccccccccccccccccccccccccc}
O_1&O_2&O_3&O_4&O_5&O_6&P_1&P_2&P_3&P_4&P_5&P_6&P_7&P_8&P_9&P_{10}&P_{11}

&P_{12}&R_1&R_2&R_3&R_4&R_5&R_6\\
-2&6&0&0&0&0&0&0&0&0&0&0&0&0&0&0&0&0&0&0&0&0&0&0\\
6&-2&-\frac{2}{9}&\frac{2}{3}&-\frac{2}{9}&\frac{2}{3}&0&0&0&0&0&0&0&0&0&0&0&0&0&0&0&0&0&0\\

0&0&-\frac{22}{9}&\frac{22}{3}&-\frac{4}{9}&\frac{4}{3}&0&0&0&0&0&0&0&0&0&0&0&0&0&0&0&0&0&0\\

0&0&\frac{44}{9}&\frac{4}{3}&-\frac{10}{9}&\frac{10}{3}&0&0&0&0&0&0&0&0&0&0&0&0&0&0&0&0&0&0\\

0&0&0&0&2&-6&0&0&0&0&0&0&0&0&0&0&0&0&0&0&0&0&0&0\\
0&0&-\frac{10}{9}&\frac{10}{3}&-\frac{10}{9}&-\frac{38}{3}&0&0&0&0&0&0&0&0&0&0&0&0&0&0&0&0&0&0\\

0&0&0&0&0&0&-16&0&0&0&0&0&0&0&-\frac{2}{9}&\frac{2}{3}&-\frac{2}{9}&\frac{2}{3}&

0&0&0&0&0&0\\
0&0&0&0&0&0&0&-16&0&0&0&0&0&0&-\frac{2}{9}&\frac{2}{3}&-\frac{2}{9}&\frac{2}{3}&

0&0&0&0&0&0\\
0&0&0&0&0&0&0&0&-16&0&0&0&0&0&-\frac{2}{9}&\frac{2}{3}&-\frac{2}{9}&\frac{2}{3}&

0&0&0&0&0&0\\
0&0&0&0&0&0&0&0&0&-16&0&0&0&0&-\frac{2}{9}&\frac{2}{3}&-\frac{2}{9}&\frac{2}{3}&

0&0&0&0&0&0\\
0&0&0&0&0&0&0&0&0&0&-16&0&0&0&-\frac{2}{9}&\frac{2}{3}&-\frac{2}{9}&\frac{2}{3}&

0&0&0&0&0&0\\
0&0&-\frac{2}{9}&\frac{2}{3}&-\frac{2}{9}&\frac{2}{3}&0&0&0&0&0&-16
&0&0&0&0&0&0&0&0&0&0&0&0\\ 
0&0&-\frac{2}{9}&\frac{2}{3}&-\frac{2}{9}&\frac{2}{3}&0&0&0&0&0&0
&-16&0&0&0&0&0&0&0&0&0&0&0\\
0&0&-\frac{2}{9}&\frac{2}{3}&-\frac{2}{9}&\frac{2}{3}&0&0&0&0&0&0
&0&-16&0&0&0&0&0&0&0&0&0&0\\
0&0&0&0&0&0&0&0&0&0&0&0&0&0&-\frac{22}{9}&\frac{22}{3}&-\frac{4}{9}&\frac{4}{3}

&0&0&0&0&0&0\\
0&0&0&0&0&0&0&0&0&0&0&0&0&0&\frac{44}{9}&\frac{4}{3}&-\frac{10}{9}&\frac{10}{3}

&0&0&0&0&0&0\\
0&0&0&0&0&0&0&0&0&0&0&0&0&0&0&0&2&-6&0&0&0&0&0&0\\
0&0&0&0&0&0&0&0&0&0&0&0&0&0&-\frac{10}{9}&\frac{10}{3}&-\frac{10}{9}&-\frac{38}{3}

&0&0&0&0&0&0\\
0&0&0&0&0&0&0&0&0&0&0&0&0&0&0&0&0&0&-2&6&0&0&0&0\\
0&0&0&0&0&0&0&0&0&0&0&0&0&0&-\frac{2}{9}&\frac{2}{3}&-\frac{2}{9}&\frac{2}{3}

&6&-2&0&0&0&0\\ 
0&0&0&0&0&0&0&0&0&0&0&0&0&0&0&0&0&0&0&0&-2&6&0&0\\
0&0&0&0&0&0&0&0&0&0&0&0&0&0&-\frac{2}{9}&\frac{2}{3}&-\frac{2}{9}&\frac{2}{3}

&0&0&6&-2&0&0\\ 
0&0&0&0&0&0&0&0&0&0&0&0&0&0&0&0&0&0&0&0&0&0&-2&6\\
0&0&0&0&0&0&0&0&0&0&0&0&0&0&-\frac{2}{9}&\frac{2}{3}&-\frac{2}{9}&\frac{2}{3}

&0&0&0&0&6&-2\\ 
\end{array}
\right)
}
\eeq

\linespread{1.3}
The $4\times 4$-block of the magnetic penguins looks like

\beq
\renewcommand{\arraystretch}{1.5}
\left(\begin{array}{cccc}
O_7&O_8&\tilde O_7&\tilde O_8\\
\frac{32}{3}&0&0&0\\
-\frac{32}{9}&\frac{28}{3}&0&0\\
0&0&\frac{32}{3}&0\\
0&0&-\frac{32}{9}&\frac{28}{3}
\end{array}
\right)
\eeq
\\
The four last columns that mix four-fermion operators with magnetic 
penguins are
given as rows. They are
\beq
\renewcommand{\arraystretch}{1.5}
\setlength{\arraycolsep}{0.04cm}
\left(
\begin{array}{lcccccccccccccc}
&O_1&O_2&O_3&O_4&O_5&O_6&P_1&P_2&P_3&P_4&P_5&P_6\\
O_7&0&\frac{832}{81}&-\frac{928}{81}&\frac{272}{81}&\frac{64}{9}&-\frac{592}{81}

&0&0&0&0&0&\frac{400}{81}\\
O_8&6&\frac{140}{27}&\frac{1090}{27}&\frac{1024}{27}&-\frac{118}{3}&
-\frac{1406}{27}&0&0&0&0&0&-\frac{238}{27}\\
\tilde O_7&0&0&0&0&0&0&-\frac{896}{81}&-\frac{896}{81}&\frac{400}{81}
&\frac{400}{81}&\frac{400}{81}&0\\
\tilde O_8&0&0&0&0&0&0&-\frac{238}{27}&-\frac{238}{27}&-\frac{238}{27}
&-\frac{238}{27}&-\frac{454}{27}&0\\
\\
&P_7&P_8&P_9&P_{10}&P_{11}&P_{12}&R_1&R_2&R_3&R_4&R_5&R_6\\
&\frac{400}{81}&\frac{400}{81}&0&0&0&0&0&0&0&0&0&0\\
&-\frac{238}{27}&-\frac{454}{27}&0&0&0&0&0&0&0&0&0&0\\
&0&0&-\frac{928}{81}&\frac{272}{81}&\frac{64}{9}&-\frac{592}{81}&0
&\frac{832}{81}&0&\frac{832}{81}&0&-\frac{464}{81}\\
&0&0&\frac{1090}{27}&\frac{1024}{27}&-\frac{118}{3}&-\frac{1406}{27}&6
&\frac{140}{27}&6&\frac{140}{27}&6&\frac{140}{27}
\end{array}
\right)
\eeq
\\
We emphasize that the matrix depicted here is $\gamma^{\mathrm{eff}}$. 
In the
HV-scheme it should coincide with the uncorrected $\gamma$. We have 
checked this
explicitly for all the entries.

\section{Definition of the functions $F_1$ - $F_4$} \label{af}

\begin{figure}[bp]

\begin{center}
\epsfbox{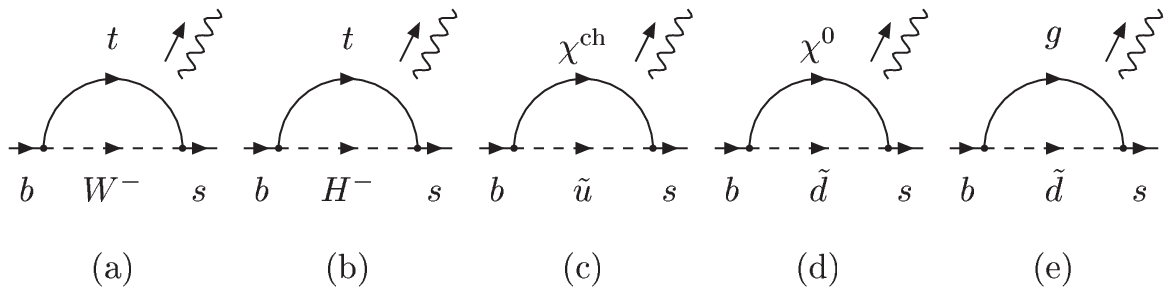}
\end{center}
\caption{\label{fig:6} Diagrams that contribute to the matching of 
$O_{7/8}$
and $\tilde O_{7/8}$. The outgoing photon/gluon is attached at every 
possible
position.}
\end{figure}

These functions appear in all calculations of diagrams like those of 
Fig.\ \ref{fig:6}.

\begin{eqnarray}
F_1(x)&=&\frac{1}{12(1-x)^4}\left[2+3x-6x^2+x^3+6x\log x\right] \nonumber \\
F_2(x)&=&\frac{1}{12(1-x)^4}\left[1-6x+3x^2+2x^3-6x^2\log x\right] \nonumber \\
F_3(x)&=&\frac{1}{2(1-x)^3}\left[-3+4x-x^2-2\log x\right] \nonumber \\
F_4(x)&=&\frac{1}{2(1-x)^3}\left[1-x^2+2x\log x\right]
\end{eqnarray}


\begin{table}[tbp]
\caption{\label{tbl:MSSM} Superfields of the MSSM with their component fields}
    \centering
    \begin{tabular}{lccc}
	&\bf{Superfield} & \bf{Boson} & \bf{Fermion} \\
	\hline 
	(S)quarks & $Q_{i}$ 
	& $\tilde{q}_{i}=\left(\begin{array}{l}\tilde{u}_{Li}\\ 
	\tilde{d}_{Li} \end{array}\right)$ & $q_{i}=\left(\begin{array}{l} 
	u_{Li}\\ d_{Li} \end{array}\right)$\\
	& $U^{c}_{i}$ & $\tilde{u}^{\dagger}_{Ri}$ & $u^{c}_{Ri}$ \\
	& $D^{c}_{i}$ & $\tilde{d}^{\dagger}_{Ri}$ & $d^{c}_{Ri}$ \\
	\hline
	(S)leptons & $L_{i}$ & 
	$\tilde{\ell}_{i}=\left(\begin{array}{l}\tilde{e}_{Li}\\ 
	\tilde{\nu}_{Li} \end{array}\right)$ & 
$\ell_{i}=\left(\begin{array}{l} 
	e_{Li}\\ \nu_{Li} \end{array}\right)$ \\
	& $e^{c}_{i}$ & $\tilde{e}_{i}^{\dagger}$ & $e^{c}_{Ri}$ \\
	\hline
	Higgs(inos) & $H_{1}$ & 
	$\tilde{h}_{1}=\left(\begin{array}{l}\tilde{h}_{1}^{0}\\ 
	\tilde{h}_{1}^{-} \end{array}\right)$ & 
$h_{1}=\left(\begin{array}{l} 
	h_{1}^{0}\\ h_{1}^{-} \end{array}\right)$ \\
	& $H_{2}$ & $\tilde{h}_{2}=\left(\begin{array}{l}\tilde{h}_{2}^{+}\\ 
	\tilde{h}_{2}^{0} \end{array}\right)$ & 
$h_{2}=\left(\begin{array}{l} 
	h_{2}^{+}\\ h_{2}^{0} \end{array}\right)$ \\
	\hline
	Gauge fields & $V^{a}$ & $A^{a}_{\mu}$ & $\lambda^{a}$\\
	\end{tabular}
\end{table}

\end{document}